\documentclass{aa}
\usepackage{psfig}
\usepackage{amssymb}
\begin{document}
	\thesaurus{08    % A&A Section
              (02.13.1;  % MHD,
               08.06.2;  % Stars: formation,
               08.13.2;  % Stars: mass-loss,
               09.10.1)} % ISM:jets and outflows.
	\title{Magnetized Protostellar Bipolar Outflows}
	\subtitle{I. Self-similar model}
	\author{Thibaut Lery $^{1}$, 
	Richard N. Henriksen $^{1}$, 
	Jason Fiege $^{2}$}
	\offprints{T. lery}
	\mail{lery@astro.queensu.ca}
	\institute {$^{1}$
	   Department of Physics,
	   Queen's University, Kingston,
	   Ontario, K7L 3N6, Canada
	\and $^{2}$
           Dept. of Physics and Astronomy,
           McMaster University,
           1280 Main St. W.,
           Hamilton, Ontario,
           L8S 4M1, Canada}
\titlerunning{Magnetized Bipolar Outflows. I}
\authorrunning{Lery et al.}
	\maketitle
	\begin{abstract}

We study a self-similar circulation model for protostellar bipolar outflows. 
The model is axisymmetric and stationary, and now includes Poynting flux. 
Compared to an earlier version of the model, this addition produces 
faster and more collimated outflows. Moreover the luminosity needed for the 
radiative heating is smaller. The solutions are developed within the context 
of $r$-self-similarity, which is a separated type of solution wherein a power
 of $r$ multiplies an unknown function of $\theta$. For outflows surrounding
 a fixed point mass the velocity, density and magnetic field respectively 
scale with spherical radius $r$ as $\vec v \propto r^{-1/2}$, $\rho \propto 
r^{2\alpha-1/2}$ and $\vec B \propto r^{\alpha-3/4}$. The parameter $\alpha$ 
must be larger than $-1/2$ and smaller than or equal to $1/4$. We obtain the 
$\theta$-dependence of all flow quantities. The solutions are characterized 
by $\alpha$ together with the scaled  temperature parameter $\Theta$ that 
imposes the necessary heat transfer. The model has been applied to both low 
and high mass protostars. Monte Carlo methods have been used to explore 
systematically the parameter space. An inflow/outflow pattern including 
collimation of high speed material and an infalling toroidal disc arises 
naturally. The disc shape depends on the imposed heating, but it is 
naturally Keplerian given the central point mass. Outflows can have large 
opening angles, that increase when magnetic field weakens. Massive 
protostars produce faster but less collimated outflows than less massive 
protostars. The model is now at a stage where synthetic CO spectra 
reproduce very well the observational features. The results strengthen 
the idea that radiative heating and  Poynting flux are ultimately the 
energy sources driving the outflow.

\end{abstract}

\section{Introduction}

General outflows and  collimated jets are thought to be intimately related 
to infall and protostellar accretion. Recent observations 
(Donati et al. \cite{donal97}, Guenther \& Emerson \cite{gue96}) 
suggest the relevance of magnetic fields to star formation, first suggested 
by Mestel \& Spitzer (\cite{ms56}). Tomisaka (\cite{tomisaka}) has studied 
numerically the dynamical collapse of magnetized molecular cloud cores from 
the runaway cloud collapse phase to the central point mass accretion phase. 
He finds that the evolution of the cloud contracting under its self-gravity 
is well expressed by a self-similar solution. Moreover inflow-outflow 
circulation appears as a natural consequence of the initial configuration.
His results support the magnetized self-similar models as recently 
presented by Fiege and Henriksen (\cite{fiege1} a,b, hereafter 
FH1, FH2) following Henriksen (\cite{henrik}, \cite{henrik2}), and 
Henriksen and Valls-Gabaud (\cite{HVG}, hereafter HV). In those models, 
the self-similar gravitationally driven convective circulation in a heated 
quadrupolar protostar envelope is solved rigorously. Moreover recent 
observations (Greaves \& Holland \cite{GH}) of the polarized dust emission 
from six star-forming cloud cores have revealed, for the first time, some 
very large twists in the magnetic field. This is 
consistent with FH1 and HV where circumstellar magnetic fields follow 
large-scale streamlines. This remains very nearly the case in the present 
model.

The self-similar models regard the molecular outflow as a natural 
consequence of the circulation established by the collapse of the 
pre-stellar cloud. {\em The models describe an outflow velocity that increases 
toward the axis of rotation, a convective pattern of infall/outflow, 
self-consistent axial collimating magnetic fields and rotation, and 
``cored apple'' type distributions of circum-protostellar gas} (Andre et al. 
\cite{andre}).  

Such models may imply a natural connection between the fast ionized jets
seen near the polar axis of the wind, and the slower and less-collimated 
molecular outflows that surround them. However velocities obtained by FH1 
and HV are smaller than some  observed velocities (e.g. EHV outflow 
($V\approx 100~km~s^{-1}$), Bachiller et al. \cite{bachiller}) unless
the solution is pushed very close to the central mass where the limiting 
speed  is about the escape speed (and the jet speed) of a few hundred 
$km~s^{-1}$. But this is precisely the region where the inner boundary 
layer may be dominant and the assumptions of this type of model are liable 
to break down. Moreover the luminosity needed for radiative heating in FH1
is too large in order to drive high-velocity outflows if only dust opacity is
used. In the present work we show that {\em the Poynting flux increases both 
the velocity and collimation of the modeled outflows by helping to transport
mass and energy from the equatorial regions}. This is much as has been argued 
for some time by other authors (e.g. Pudritz and Norman \cite{PN}, Ouyed 
and Pudritz \cite{ouyedp}), but our models are globally consistent in 
space. Self-similar models can not of course be globally consistent in 
time even if they are not stationary (we have found such models and will 
report them elsewhere), since they are ignorant of initial conditions. And 
in fact they must also be `intermediate' in space since there 
are inevitably small regions near the equator, the axis and at the centre 
which are excluded from the domain of self-similarity. Thus, we derive 
globally self-consistent models of bipolar outflows and infall accretion, but
specifically exclude the regions dominated by the disc, jet, and protostar.

Qualitatively we may understand the quadrupolar global circulation in the 
following fashion. As material falls towards the central object, 
it is gradually slowed down by increasing radial pressure gradients
due to the rising temperature, density, and magnetic field encountered 
by material near the central object. This pressure barrier, with the help 
of centrifugal forces, deflects and accelerates much of the infalling 
matter into an outflow along the axis of symmetry. The outflow 
velocity is naturally the highest near the axis of rotation since the 
pressure gradients are strongest there. Mathematically the quadrupolar model 
can be seen as a form of instability of the spherical Bondi accretion flow, 
wherein rotation, magnetic fields and anisotropic heating transform a central 
nodal singularity into a saddle point singularity.

The pressure gradients also act to 
collimate the flow, forming consistently the now traditional de Laval 
nozzle (e.g. Blandford and Rees \cite{br}, K\" onigl \cite{ko1}).
We find however that the most convincing flows are everywhere 
super-Alfv\'enic, in agreement with FH1, but the nozzle may still be 
critical in terms of the fast magnetosonic speed as discussed in 
FH1 and below. We note that one can expect conical shocks near the 
rotational axis, which would seem to be necessary to explain the 
shocked molecular hydrogen emission often observed.  

The paper is organized as follows:
in \S2 we introduce the model and its approximations, and discuss their 
consequences; numerical solutions are presented in \S3 for 
virial-isothermal and radiative models describing outflows from low
and high mass protostars. In \S4 we study the ensuing synthetic radio 
maps; this section is followed by discussions of the results in \S5, 
and conclusions are presented in \S6.

%%%%%%%%%%%%%%%%%%%%%%%%%%%%%%%%%%%%%%%%%%%%%%%%%%%%%%%%%%%%%
\section{The Analytical Model}
%%%%%%%%%%%%%%%%%%%%%%%%%%%%%%%%%%%%%%%%%%%%%%%%%%%%%%%%%%%%%

The main difference between hydrodynamic and MHD outflows is the low 
asymptotic Mach number in the MHD case. For protostellar outflows, the
magnetosonic speed is about $100~km~s^{-1}$ at $10^5$ AU and therefore 
far exceeds the sonic speed. It is not a priori obvious that the internal 
Alfv\'enic Mach number of the jet or outflow is small. However most 
of our models are super-Alfv\'enic everywhere.

The plasma is supposed to be perfectly conducting and 
therefore the flow is governed by the basic equations of ideal MHD 
without taking into account resistivity or ambipolar diffusion. 
In order to make the system tractable we assume axisymmetric flow 
so that $\partial/\partial\phi=0$ and all flow variables are functions 
only of $r$ and $\theta$. 

Kudoh and Shibata (\cite{kudoh2}) have performed time dependent 
one-dimensional (1.5-dimensional) magnetohydrodynamic numerical 
simulations of astrophysical jets that are magnetically driven 
from Keplerian discs. They have found that the jets, which are 
ejected from the disc, have the same properties as the steady 
magnetically driven jets they had found before (Kudoh \& Shibata 
\cite{kudoh1}). Their numerical results suggest that a steady 
model, as we are assuming in the present work (i.e. 
$\partial/\partial t=0$), is a good first approximation to the 
time-dependent problem on time-scales short compared to that 
required to dissipate the circumstellar material. 
Time independence will be relaxed in future work.  

Accretion discs near protostars are probably heavily convective and 
therefore prone to dynamo action (Brandenburg et al. \cite{brand}, FH1). 
Since the ensuing magnetic fields are loaded with disc plasma, 
currents are allowed to circulate between the disc surface and 
the wind region above the disc. For rapidly rotating discs, this 
dynamical system may evolve naturally into a quadrupolar structure 
(Camenzind \cite{cam}, Khanna and Camenzind \cite{kc}). The 
strong differential rotation of accretion discs is responsible for 
the excitation of quadrupolar modes in discs. Consequently, in the 
present model, magnetic field and streamlines are required to be 
quadrupolar in the poloidal plane. A moderate 
amount of heating supplied near the turning point allows the outflow 
to possess finite velocities at infinity.

These considerations have led us to use the set of equations of steady, 
axisymmetric, ideal MHD and to seek solutions with a quadrupolar 
geometry. Even though the poloidal components of the 
magnetic and velocity fields have to be parallel, the toroidal 
components need not  share the same constant of proportionality 
(Henriksen \cite{henrik3}). This permits a poloidal  conservative electric 
field to exist in the inertial frame, and so admits steady 
Poynting flux driving. One needs then to introduce an electric 
potential function and an azimuthal magnetic field independent of the 
azimuthal velocity. This point is the major improvement of the model 
with respect to FH1. In the radiative case, the equations describing 
diffusive radiative transfer (to zero order in $v/c$) and a 
Kramer's-type law for the Rosseland opacity are added as in FH1. 
We also neglect the self-gravity of the protostellar material by 
assuming that the gravitational field is dominated by the central 
mass that is assumed to be an external fixed parameter in our model. 
Henceforth the model only applies to a state where the protostellar 
core is already formed.

It is unlikely that this model can be extended to include the optical jets 
since these must originate very near to the central object where the model 
may not apply in its steady form. Nevertheless, the inclusion of a jet 
model in the axial region, such as given by lery et al.(\cite{lery1},
\cite{lery2}), could remedy this lack in order to make a more global model 
for protostellar objects.

%%%%%%%%%%%%%%%%%%%%%%%%%%%%%%%%%%%%%%%%%%%%%%%%%%%%%%%%%%%%%
\subsection{Scales and Self-similar Laws}
%%%%%%%%%%%%%%%%%%%%%%%%%%%%%%%%%%%%%%%%%%%%%%%%%%%%%%%%%%%%%

Our model is developed within the context of  
power law self-similar models as developed by Henriksen (\cite{henrik}), 
and in HV and FH1. A model with such symmetry was already 
used by Bardeen \& Berger (\cite{BB}) for galactic winds, although the 
present development has been independent. Parallel developments have 
been largely occupied with the study of winds from an established 
accretion disc (Blandford \& Payne \cite{bp}, Konigl \cite{ko}, 
Ferreira \& Pelletier \cite{ferr1}, Rosso \& Pelletier \cite{RossoP}, 
Ferreira \cite{ferr2}, Contopoulos \& Lovelace \cite{conto}, Contopoulos 
\cite{contop}, Sauty \& Tsinganos \cite{sautyt}, Tsinganos et al. 
\cite{tsinetal}, Vlahakis \& Tsinganos \cite{VT}). Our use of this 
model in order to study inflow and outflow as part of the global 
circulation dynamics around protostars seems unique. The form used 
corresponds to an example of `incomplete self-similarity' in the 
classification of Barenblatt (\cite{baren}), but fits into the 
general scheme of self-similarity advocated by Carter and Henriksen 
(\cite{CH}) with $r$ as the direction of the Lie self-similar motion.
We work in spherical polar coordinates $r, \theta, \phi$ centered 
on the point mass $M$ and having the polar axis directed along the 
mean angular momentum of the surrounding material. 

The self-similar symmetry is identical to that used in FH1 except that 
two quantities $y_{\phi}$ and $y_{p}$ defined such that  $\sqrt{4\pi\mu}
y_{\beta}\equiv v_{\beta}/(B_{\beta}/\sqrt{4\pi\rho})$ (where $\beta$ 
indicates the appropriate component) replace $y$ in FH1. The power laws 
of the self-similar symmetry are determined, up to a single parameter 
$\alpha$, if we assume that the local gravitational field is dominated 
by a fixed central mass. In terms of a fiducial radial distance, $r_o$, 
the self-similar symmetry is sought as a function of two scale invariants, 
$r/r_o$ and $\theta$, in a separated power-law form. The equations of 
radiative MHD require the following radial scaling relations for the 
variables:
\begin{equation}
{\bf v} = \sqrt{\frac{GM}{r_o}} \left(\frac{r}{r_o}\right)^{-1/2} 
\ {\bf u}(\theta),
\label{self1}
\end{equation}
\begin{equation}
B_{\phi} = \sqrt{\frac{G M^2}{r_o^4}} \left(\frac{r}{r_o}\right)
^{\alpha-3/4} \frac{u_{\phi}(\theta)}{y_{\phi}(\theta)},
\end{equation}
\begin{equation}
B_{p} = \sqrt{\frac{G M^2}{r_o^4}} \left(\frac{r}{r_o}\right)
^{\alpha-3/4} \frac{u_{p}(\theta)}{y_{p}(\theta)},
\end{equation}
\begin{equation}
\rho = \frac{M}{r_o^3} \left(\frac{r}{r_o}\right)
^{2\alpha-1/2} \mu(\theta),
\end{equation}
\begin{equation}
p = \frac{G M^2}{r_o^4} \left(\frac{r}{r_o}\right)
^{2\alpha-3/2} P(\theta),
\end{equation}
\begin{equation}
\frac{k T}{\bar\mu m_H} = 
\frac{G M}{r_o} \left(\frac{r}{r_o}\right)^{-1}
\ \Theta(\theta),
\label{self2}
\end{equation}
\begin{equation}
\vec F_{rad}=\left(\frac{GM}{r_o}\right)^{3/2}
\frac{M}{r_o^3}\left(\frac{r}{r_o}\right)^{\alpha_f-2} \vec f(\theta) .
\label{selfrad}
\end{equation}
In these  equations the microscopic constants are represented by
$k$ for Boltzmann's constant, $\bar\mu$ for the mean atomic weight,
$m_H$ for the mass of the hydrogen atom. The self-similar index $\alpha$ 
is imposed as a parameter of the solution, but as in FH1 it is constrained to 
lie in the range $1/4\ge \alpha> -1/2$. In the last equation, the index
$\alpha_f$ is a measure of the loss (if negative) or gain in radiation energy
as a function of radial distance.

Whenever we take account of radiative diffusion explicitly, as well as for 
the definition of the luminosity,   we will use the 
same formulation as in FH1. The simplest approach however is to assume that 
the temperature is some fraction of the `virialized' value. That is, according 
to relation (\ref{self2}), that $\Theta$ takes some constant value $\le 1$. 
By using the first law and assuming an ideal gas (so that with the above 
relations for $p$ and $\rho$ we have $P(\theta)=\mu \Theta$ ) this 
assumption implies that on a sphere the specific entropy $s$ varies 
according to 
\begin{equation}
Tds=-\frac{kT}{\bar\mu m_H} d\ln\mu.
\label{anglentropy}
\end{equation} 
Consequently we are explicitly adding or subtracting heat from the system 
on a spherical surface according to the sign of $-d\ln\mu$. In our 
models this normally increases towards the rotational axis. Thus, at least 
implicitly,  we {\it always} employ some form of heat transfer, presumably 
radiative, to the gas over the poles relative to the equatorial material. 

The same procedure for variations in the radial direction yields 
\begin{equation}
Tds=-\frac{GM}{r_o} \Theta
\left[\frac{1+2(\gamma-1)(\alpha-1/4)}{\gamma-1}\right]
\left(\frac{r_o}{r}\right)^2 \frac{dr}{r_o}.
\label{radentropy}
\end{equation}
Energy is therefore lost to the material with increasing radius provided 
that the quantity in the square bracket is positive. This requires 
$\alpha-1/4>-\frac{1}{2(\gamma-1)}$, which is true for reasonable ratios 
of specific heats throughout our permitted range of $\alpha$.

In order for radiative heating to be adequate we must have 
$\tau \vec F_{rad}/c$ sufficiently large, where $\vec F_{rad}$ 
is the net radiative flux through a mass element. We note
that because of the dependence on $\vec F_{rad}$, this is not the same  
as requiring an optical depth of order unity. Nevertheless in order for 
the radiative transfer by diffusion to be plausible, this latter condition 
must also be satisfied. 

In either case the issue depends on the nature of the opacity $\kappa$. We 
follow FH1 in supposing that this is predominantly a dust opacity 
(at least at some distance from the `star') which is taken in Kramer's form 
\begin{equation} 
\kappa=\kappa_o \rho^a T^b.
\label{opacity}
\end{equation}
A fit to a dust model in FH1 yields $a=0$, and $b\approx 2$ and $\kappa_o =
6.9\times 10^{-6} cgs$. We can calculate the mean optical depth between an 
inner radius $r_*$ and the fiducial radius $r_o$, namely 
$ <\tau>\equiv \frac{1}{4\pi}\int_{r_*}^{r_o}~\kappa \rho dr\, d\Omega$, 
using this opacity and the self-similar scaling relations. We find
\begin{equation}
<\tau>=\frac{<\mu\Theta^2>}{|2\alpha-3/2|}\left[
\left(r_*/r_o\right)
^{(2\alpha-3/2)}-1\right],
\label{optdepth}
\end{equation}
where we have defined the fiducial radius as
\begin{equation}
r_o\equiv \left[\kappa_o\left(\frac{\bar{\mu}m_H}{k}\right)^2 
G^2\right]^{1/4} M^{3/4},
\label{rfid}
\end{equation}
which is numerically
\begin{equation}
r_o\approx 1400 (M/M_\odot)^{3/4} AU.
\label{ro}
\end{equation}
Thus {\em we obtain a fiducial radius that is characteristic of the bipolar 
outflow sources} and is a convenient unit for our discussions.  A similar 
scale (to within a factor 5) was previously deduced in Henriksen 
(\cite{henrik}) based on the source luminosity and 
fundamental constants (including the radiation constant). We note that 
the external opacity from $\infty$ to $r_o$ is given simply by 
$ <\tau>_{ext}=<\mu\Theta^2> / |2\alpha-3/2|$ and the `photosphere' 
of the cloud is therefore found where $\tau_{ext}\approx 2/3$. 

The corresponding temperature at the fiducial radius is given by 
\begin{equation}
T\approx 164 \, \Theta \, (M/M_\odot)^{1/4}K .
\end{equation}
We observe that if we identify our radiative fiducial scale $r_o$ with an 
empirical scale $r_{em}$ from $<\mu>/r_{em}^2=2\times 10^{-9\pm 1} (AU)^{-2}$
(Richer et al. \cite{PPIV}), we can infer a relation between the mean density 
parameter $<\mu>$ and the central mass in the form 
\begin{equation}
<\mu> =3.38\times 10^{-3\pm 1}\left(M/M_\odot\right)^{3/2}.
\label{high/low}
\end{equation}
{\em This is a useful way to relate the key model parameter $\mu$ to the 
physical mass and luminosity}. We note however that the optical 
depth is not likely to approach unity for reasonable temperatures, except 
for exceptionally massive objects. But this does not prevent radiative 
heating from playing an important dynamical role near the protostar.   

In fact using $r_o$, as above, and the definition of
radiation field $\vec F_{rad}$ 
(Eq.~\ref{selfrad}), the local energy flow can be expressed by   
\begin{equation}
r^2\vec{F_{rad}}= 0.015\left(r/r_o\right)^{\alpha_f} \left(M/M_\odot
\right)^{5/8} \vec{f} L_\odot.
\end{equation}
The radiative force per unit mass is given by 
\begin{equation}
\kappa\vec{F_{rad}}/c\approx 10^{-13}\left(M/M_\odot\right)^{1/8}
\left(r/r_o\right)^{(\alpha_f-4)}\vec{f} cm~ s^{-2},
\end{equation} 
which must be less than the Eddington limit of 
$GM/r^2 =3.5\times 10^{-7}(M/M_\odot)
\left(r_o/r\right)^2 cm~ s^{-2}$, for consistency. 
These are helpful considerations for our radiative models below
that always verify this condition a posteriori.

%%%%%%%%%%%%%%%%%%%%%%%%%%%%%%%%%%%%%%%%%%%%%%%%%%%%%%%%%%%%%
\subsection{The Equations}
%%%%%%%%%%%%%%%%%%%%%%%%%%%%%%%%%%%%%%%%%%%%%%%%%%%%%%%%%%%%%

We use the self-similar forms in the usual set of ideal MHD equations 
together with the radiative diffusion equation when applicable. The 
corresponding system of equations is reported in Appendix A, and the 
ensuing first integrals in Appendix B. As noted above, in the present 
model, there is no constraint requiring parallelism between velocity 
and magnetic field as used in FH1. This allows us to discern clearly 
the effects of a non-zero Poynting flux. The self-similar variable 
directly related to magnetic field $y$ has two components that are not 
equal in the present model. These quantities are either projected in the 
poloidal plane or correspond to the toroidal components and are 
respectively defined by
\begin{equation}
y_{p,\phi}(\theta)=\frac{M_{ap,a\phi}}{\sqrt{4\pi\mu(\theta)}}.
\label{Y}\end{equation}
Consequently the system deals with two different components of the 
Alfv\'enic Mach number $M_{ap}$ and $M_{a\phi}$ defined by:
\begin{equation}
M^2_{ap,a\phi}(\theta)=\frac{\vec{v}_{p,\phi}^2}
{\vec{B}_{p,\phi}^2/4\pi\rho}.
\end{equation}

In this  model, since $r$ and $\phi$ respectively correspond to the 
directions of self-similarity and axisymmetry, only waves 
that propagate along $\theta$ in the poloidal plane can preserve 
these two symmetries. Therefore  the relevant Alfv\'en mode 
propagates in the direction $\theta$ in the poloidal plane with 
a phase velocity: 
\begin{equation}
V_{a\theta}= \frac{B_\theta}{\sqrt{4\pi \rho}}.
\end{equation}
Moreover we define an Alfv\'enic point where the total poloidal 
flow velocity is equal to this value in magnitude and direction 
(i.e. $u_r=0$), which by (\ref{Y}) is equivalent  to:
\begin{equation}
M_{ap}=1=4\pi\mu y_p^2.
\end{equation}
One can then define a critical scaled density corresponding to the 
Alfv\'enic scaled density
\begin{equation}
\mu_a=\frac{1}{4\pi y_p^2}.
\end{equation}
The flow is super-Alfv\'enic if the density $\mu$ is 
larger than the critical Alfv\'en density $\mu_a$.
Moreover, the compressible slow and fast MHD waves propagate in a 
poloidal direction $\theta$ with  phase velocities that satisfy the 
quartic (Sauty and Tsinganos \cite{sautyt}) 
\begin{equation}
\left( V_{s/f}\right)_\theta^4-\left( V_{s/f}\right)_\theta^2
\left( V_a^2+C_s^2 \right) + C_s^2\left( V_a\right)_\theta^2=0,
 \label{dispersion}
 \end{equation}
where $C_s^2$ is the isothermal sound speed. The condition that 
$u_\theta$ be equal to one of these wave speeds (where we expect a 
singular point in the family of solutions) can be expressed in terms 
of self-similar variables as:
\begin{equation}
\left(u_\theta^2\right)_{crit}=\Theta + \frac 
{u_r^2+u_\phi^2 y_p^2 y_\phi^{-2}}{4\pi \mu y_p^2-1},
\end{equation}
and in physical units (cf FH1):
\begin{equation}
\left(v_\theta^2\right)_{crit}=\frac{P}{\rho}
\left(1-M_{ap}^{-2}\right)+ \frac{B^2}{4\pi \rho}.
\end{equation}
Equation (\ref{dispersion}) shows that whenever $V_a>>C_s$ the slow 
and fast speeds become $V_{a\theta}C_s/V_a$ and $\sqrt{V_a^2-
V_{a\theta}^2C_s^2/V_a^2}$ respectively. Thus a super-Alfv\'enic flow 
as defined above can only encounter the fast speed and then only where 
$V_a \gtrsim V_{a\theta}$. In a region where the sound speed dominates, the 
fast and slow speeds become $\sqrt{C_s^2-V_{a\theta}^2}$ and 
$V_{a\theta}$ respectively. Either the sound speed or the Alfv\'en 
speed is dominant according as $\Theta$ is larger or smaller than 
$\vec{u}_p^2/M_{ap}^2+\vec{u}_\phi^2/M_{a\phi}^2$.      

 As in Henriksen (\cite{henrik3}) we can consider the energy balance 
 in these models. A convenient form for the energy equation is 
  \begin{equation}
  \vec{\nabla}.\left(\rho\vec{v}(\frac{\vec{v}^2}{2}+\Phi)+\vec{S}\right)
  =-\vec{v}.\vec{\nabla}p,
  \label{energy}
  \end{equation}
  where the Poynting flux vector is 
  \begin{equation}
  \vec{S}=\frac{\vec{B}^2}{4\pi}\vec{v}-\left(\frac{\vec{B}.
  \vec{v}}{4\pi}\right) \vec{B},
  \end{equation}
  and $\Phi=-GM/r$. In self-similar variables the Poynting flux becomes 
\begin{equation}
\vec{S}
\equiv 
\left(\frac{G M}{r_o}\right)^{\frac{3}{2}}
\left(\frac{M}{r_o^3}\right)\left(\frac{r}{r_o}\right)
^{2\alpha-2}\vec{\sigma}
(\theta)
\label{poynting}
\end{equation} 
where the self-similar Poynting flux $\vec{\sigma}(\theta)$ 
can be defined in terms of the other variables as:
\begin{equation}
\vec{\sigma}(\theta)=
\left(\frac{1}{y_p}-\frac{1}{y_\phi}\right)
\left(\vec{u_\phi} \frac{u_p^2}{y_p}-
\vec{u_p}\frac{u_\phi^2}{y_\phi}\right),
\label{poynting2}
\end{equation}
 The energy equation becomes
 \begin{eqnarray}
 &2\alpha\left[\mu u_r(\frac{\vec{u}^2}{2}-1)+\sigma_r\right]
 +(2\alpha-{\frac{3}{2}})u_r P
 \nonumber\\
 &= 
-\frac{1}{\sin{\theta}}
 \frac{d}{d\theta}\left[\sin{\theta}(\mu u_\theta(\frac{\vec{u}^2}{2}-1)+
 \sigma_\theta)\right]
-\frac{u_\theta}{\sin{\theta}}
\frac{dP}{d\theta}.
 \end{eqnarray}
 This equation is generally true for our models but it does not provide 
 an independent integral for non-zero pressure. This we treat in the next 
 sub-section.
%%%%%%%%%%%%%%%%%%%%%%%%%%%%%%%%%%%%%%%%%%%%%%%%%%%%%%%%%%%%%
\subsection{The Zero Pressure Limit}
%%%%%%%%%%%%%%%%%%%%%%%%%%%%%%%%%%%%%%%%%%%%%%%%%%%%%%%%%%%%%

In the zero pressure limit one can use the self-similar energy equation  
 together with $\sigma_r=(u_r/u_\theta)\sigma_\theta$ and 
 $u_r/u_\theta=d\ln{r}/d\theta$ on a stream line to write a Bernoulli 
 integral 
 \begin{equation}
 r^{2\alpha}\sin{\theta} 
\left[ \mu u_\theta(\vec{u}^2/2-1)+\sigma_\theta \right]=B,
 \end{equation}
 where $B$ is the Bernoulli constant on each stream line. In view of 
 Eq.~\ref{stream1} we can write this as 
 \begin{equation}
 \frac{B}{\eta_1}=
\frac{1}{r(\theta)}\left[
\vec{u}^2/2-1
+\frac{u_\phi^2}{4\pi\mu y_\phi}
\left(\frac{1}{y_\phi}-\frac{1}{y_p}\right)\right], \label{Bernoulli} 
\end{equation} 
where the second term evidently describes the Poynting driving 
(magnetic energy term) and the first term gives the kinetic and 
gravitational energies. Hence if $B\ne 0$, we require the expression 
contained in brackets on the RHS of this last equation to tend towards 
infinity both at the equator and on the polar axis since 
$r(\theta)\rightarrow \infty$ in these limits. This is an improbable outer 
boundary condition  although physically realizable at some large but 
finite radius (which might be a function of angle and so trace out an 
excluded region for the self-similar solution). It does 
permit the exchange between say a large driving Poynting term near the 
equator and a large kinetic energy term near the axis, but the source of 
the `free' magnetic energy is left undetermined. 

We turn to the special case $B=0$, which requires this same factor to 
be zero everywhere on a streamline. In order that there be an exchange 
between the kinetic term and  the Poynting term in the sum  we must have 
either $\vec{u}^2<2$ and $y_\phi<y_p$ or the reverse of both inequalities. 
In the first case we do not produce velocities greater than the escape 
velocity anywhere, while in the second case we require $M_{ap}<M_{a\phi}$ 
everywhere on the streamline which also seems improbable. We are left 
then with the first possibility, which includes the case where each term 
vanishes separately on the boundaries. This suggests that starting from 
a physically likely free-fall boundary condition, $v$ is less than or 
equal to the escape velocity on every stream line in the absence of 
pressure, becoming equal again to the escape velocity at large distances. 

Thus without pressure effects one is unable to produce net outflows at 
infinity without appealing to a Poynting flux from an excluded equatorial 
region. However one does thus deflect and collimate the material 
(magnetically and centrifugally) in a lossless way, and so  all energy 
added near the star may appear in the outflow at infinity rather than be 
lost to work against gravity. The fact that there is no net gain of kinetic 
energy from the magnetic field energy on a stream line is peculiar  to the
 case with $B=0$. It is  due to the fact that energy that is first gained 
by the magnetic field at the expense of gravity and rotation during infall 
is subsequently returned as gravitational potential 
energy and kinetic energy during the outflow. With $B\ne 0$, there may be 
a source of Poynting flux in the excluded region (see e.g. also Henriksen
(\cite{henrik3}).     
 
%%%%%%%%%%%%%%%%%%%%%%%%%%%%%%%%%%%%%%%%%%%%%%%%%%%%%%%%%%%%%
\section{Numerical Analysis}
%%%%%%%%%%%%%%%%%%%%%%%%%%%%%%%%%%%%%%%%%%%%%%%%%%%%%%%%%%%%%

%%%%%%%%%%%%%%%%%%%%%%%%%%%%%%%%%%%%%%%%%%%%%%%%%%%%%%%%%%%%%
\subsection{The Numerical Method \label{NUM}}
%%%%%%%%%%%%%%%%%%%%%%%%%%%%%%%%%%%%%%%%%%%%%%%%%%%%%%%%%%%%%

The numerical solutions in this work are obtained by initiating 
the integrations of the system of six equations
from a chosen angle between the axis ($\theta=0$) 
and the equator ($\theta=\frac{\pi}{2}$). Starting values are 
varied until the boundary conditions are met. Solutions can not 
strictly reach the axis and therefore are limited by a minimum
angle $\theta_{min}$. Moreover $\frac{\pi}{2}$ bounds the 
solution only if we demand exact reflection symmetry about the 
equator. Solutions are thus generally defined on a wedge given by
$0<\theta_{min}\le\theta\le\theta_{max}$. In order to be able to 
start at will from either a sub- or super-Alfv\'enic point, 
a systematic search for singular points of the system has been conducted.
This allows us to restrict a first guess for the typical values to use
as input for the system. 

Several general criteria have been 
employed to further constrain the solutions. First the radial velocity 
$u_r$ must vanish once, to have inflow-outflow, and only once, 
to avoid a higher order convection pattern. We also require $u_\theta$ and 
$u_\phi$ to vanish at both boundaries, since there must be no 
mass flux across them, and since only zero angular momentum material 
can reach the axis (which all material lying strictly in the 
equator does). In practice, the first of these two conditions 
 is sufficient to impose the second one. The same constraints  
apply to the magnetic field components. We take $u_\theta$ 
to be negative in order to get a circulation pattern rising over the 
poles, and $u_\phi$ to be positive as a convention. We require also 
that $u_r$ is either decreasing or constant with increasing angle and 
that the density and pressure attain their maxima close to $\theta_{max}$.

To find solutions satisfying all of the previous requirements a 
Monte Carlo shooting method has been used starting from values
corresponding to either sub- or super-Alfv\'enic starting points. 
Solutions have been found using as starting points successively 
$\theta_{min}$, $\theta_{max}$ and  the stream-line turning point where 
the radial velocity vanishes. It has been found that the most convenient
starting point is $\theta_{min}$. Results of these investigations and 
representative cases are shown below.

%%%%%%%%%%%%%%%%%%%%%%%%%%%%%%%%%%%%%%%%%%%%%%%%%%%%%%%%%%%%%
\subsection{Low Mass Protostars \label{low:iso}}
%%%%%%%%%%%%%%%%%%%%%%%%%%%%%%%%%%%%%%%%%%%%%%%%%%%%%%%%%%%%%

%%%%%%%%%%%%%%%%%%%%%%%%%%%%%%%%%%%%
\begin{figure}
\psfig{figure=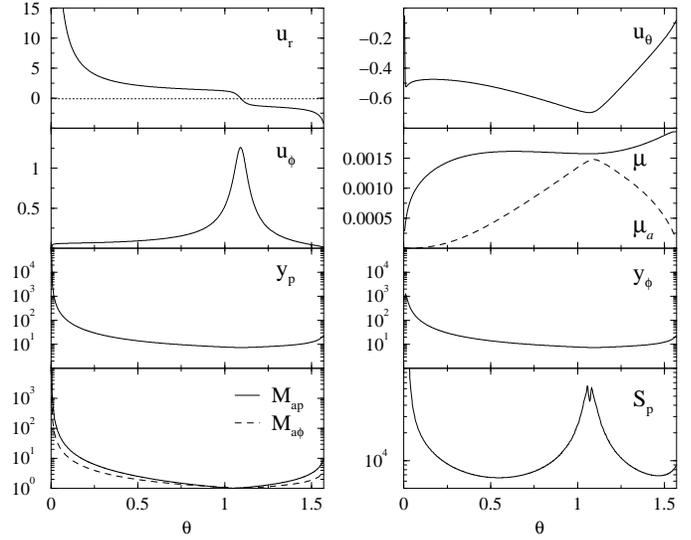,width=\linewidth}
\caption[ ]{
Variations with angle of the self-similar variables (related to 
velocity ($u_r$, $u_\theta$, $u_\phi$), 
density ($\mu$ and $\mu_a$), magnetic field ($y_p$, $y_{\phi}$)
and Mach number ($M_{ap}$, $M_{a\phi}$)
for a virial-isothermal low mass solution with 
$\alpha=-0.3$ and $\Theta=0.41$. The poloidal component $S_p$
of the Poynting flux is plotted in the lower right panel.
$M\approx 1~M_\odot$.}
\label{figsol2}
\end{figure}
%%%%%%%%%%%%%%%%%%%%%%%%%%%%%%%%

Most of the present models for star formation deal only with low mass stars. 
It is then natural to begin with this type of object. The solution shown 
in Fig.\ref{figsol2} is representative of this class in the virial-isothermal 
approximation. It shows the variations with angle of the six self-similar 
variables (related to velocity ($u_r$, $u_\theta$, $u_\phi$), density 
($\mu$ and $\mu_a$), magnetic field ($y_p$, $y_{\phi}$)) together with 
poloidal and toroidal Mach numbers ($M_{ap}$, $M_{a\phi}$) and poloidal 
component of the Poynting flux ($S_p$). In this solution we have taken 
$\alpha=-0.3$ and $\Theta=0.41$ in order to compare with solutions given 
in FH1. As discussed in the previous paragraph, various requirements have 
been met; $u_\phi$ and $u_\theta$ both vanish at 
the polar and equatorial boundaries, as do the magnetic field components 
($y$ is diverging). The maxima for rotation velocity and Poynting flux, 
corresponding to the minimum in $u_\theta$, occur where the radial 
velocity changes signs. This point defines the total opening angle of the 
outflow which in this model is of the order of $60^\circ$. The radial 
component of the Poynting flux $S_r$ changes signs as $u_r$ does, but its 
poloidal component defined by $S_p=\sqrt{S_r^2+S_\theta^2}$ remains positive 
by definition and shows only a slight dip near maximum.

The high velocity outflow is of course much more narrowly collimated than 
the general outflow ($u_r>0$). The density $\mu$ is always superior 
to the Alfv\'enic critical density and both densities are almost equal 
where $u_r$ changes sign. Thus the flow is always  super-Alfv\'enic as 
confirmed by the Mach number components in Fig.\ref{figsol2}.
Since differences between $y_p$ and $y_\phi$ are not evident to the eye in 
this figure, we have plotted the poloidal component of Poynting flux as given 
by Eq.~\ref{poynting2}. It shows the peak energy carried by the 
magnetic field to be at the turning point of the flow.
  
Thus this calculation produces wide angle relatively slow outflows 
surrounding a fast component that is identifiable with the EHV outflows 
or `molecular jets'. We shall see below that inclusion of the 
Poynting flux allows  more collimated and faster `jets' surrounded 
by a slower wide angle outflow than is the case without it. Therefore 
 the magnetic field plays a crucial role in the YSO outflow despite its 
 conservative nature.

\begin{figure*}
\psfig{figure=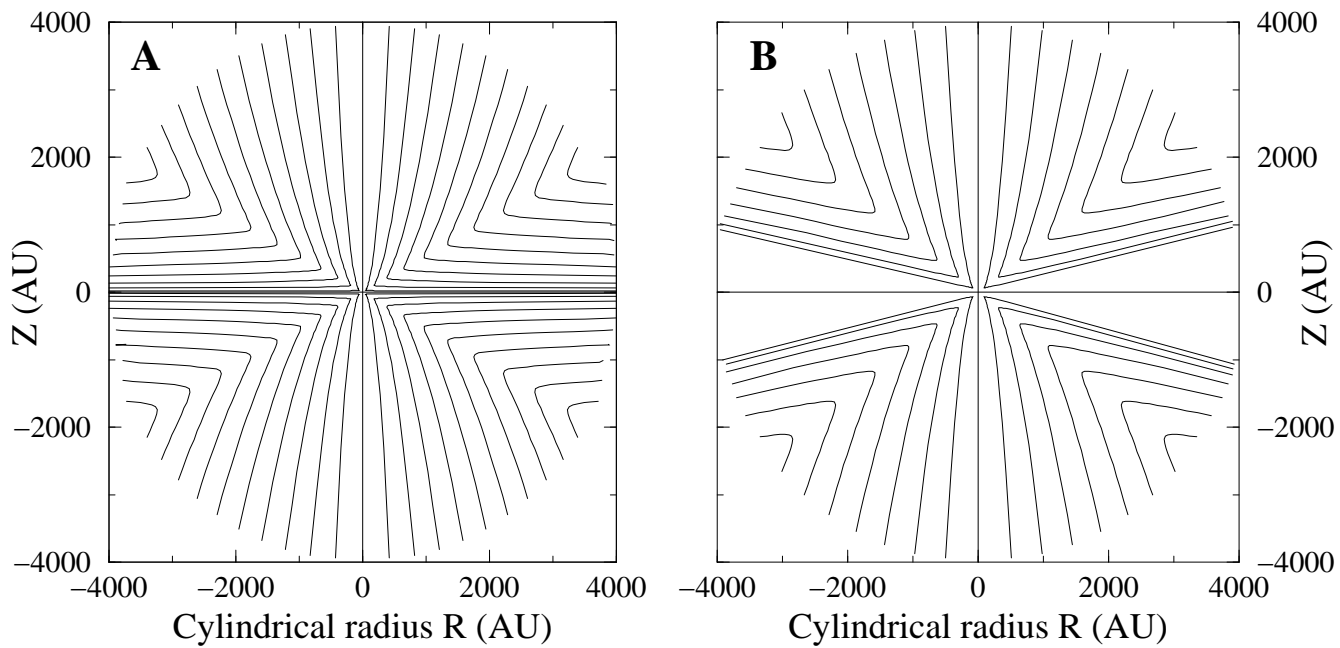,width=\linewidth}
\caption[ ]{
Streamlines of the flow in the poloidal plane for 
$\alpha=-0.3$ and $\Theta=0.41$. They are integrated 
out to a radius of 4000 AU. Panel A shows a solution 
that almost fills  full quadrant (from 
$\theta \approx 0$ to $\approx \frac{\pi}{2}$).
Solution in panel B has a smaller opening angle and 
goes from $\theta \approx 0$ to 1.33.}
\label{figSTREAM}
\end{figure*}

For embedded sources (class 0/I) most of the circumstellar matter is 
distributed in an envelope with a typical size of about $10^4$ AU (Adams 
et al. \cite{adams}, Terebey et al. \cite{tere}, Andr\'e \& Montmerle 
\cite{andreM}). The size that we will use to present our solutions will 
be of this order. For example, two poloidal projections of the streamlines 
are represented in Fig.\ref{figSTREAM} out to a radius of 4000 AU. On the 
left panel (case A), the previous solution is shown from (nearly) 
$\theta=0$ to (nearly) $\frac{\pi}{2}$. In the panel on the right 
(case B) we show the result of varying the magnetic field and density from 
the previous values. The solution shown stops at $\theta=1.33$ and 
the empty region is outside of the domain of the solution.
If the pressure and tangential velocity matching were done properly
at the boundaries, a rotating Bondi accretion flow (Henriksen and Heaton
 \cite{HenriksenHeaton}) might describe naturally a true accretion disc.
%%%%%%%%%%%%%%%%%%%%%%%%%%%%%%%%%%%%
%%%%%%%%%%%%%%%%%%%%%%%%%%%%%%%%

The self-similar structure of the streamlines does not in fact strictly
match an arbitrary boundary condition at infinity, but the calculated 
symmetry seems to be a natural small scale limit of a more general 
self-gravitating circulation flow. Moreover the maximal meridional 
velocity close to the axial region (see Fig.\ref{figsol2}) shows the 
tendency for the flow to collimate cylindrically along the axis of 
rotation. This is in agreement with Lery et al. (\cite{lery1}, \cite{lery2}) 
where it is shown that cylindrical collimation is a general asymptotic 
behavior of magnetized outflows surrounded by an external pressure, 
due in the present case to the outer part of the molecular cloud or 
the interstellar medium.

%%%%%%%%%%%%%%%%%%%%%%%%%%%%%%%%%%%%
\begin{figure}
\psfig{figure=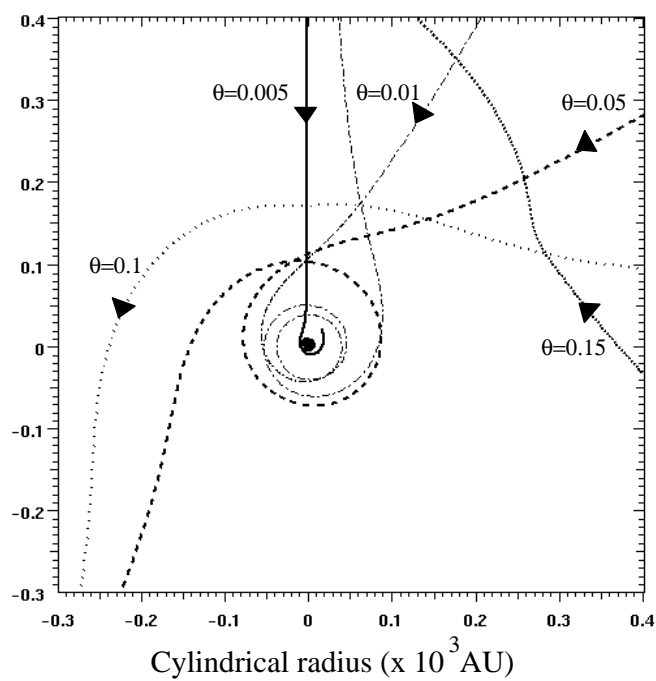,width=\linewidth}
\caption[ ]{
View down the axis of streamlines in the axial region
for the virial-isothermal solution 
($\alpha=-0.3$,$\Theta=0.41$). The streamline plotted 
with a solid line has the smallest angle with respect to 
the rotational axis. Other streamlines end with larger 
angles whose values are given for each streamline.}
\label{fig:above}
\end{figure}
%%%%%%%%%%%%%%%%%%%%%%%%%%%%%%%%

Fig.\ref{fig:above} presents a zoom of streamlines in the axial region 
view from above . Streamlines start in the equatorial region and end in 
the axial region with small angles ($\theta=0.005$ to 0.15). In addition 
to the meridional behavior shown previously in Fig.\ref{figSTREAM}, each 
streamline makes a spiraling approach to the axis and then emerges in the 
form of an helix wrapped about the axis of symmetry. The closer in angle to 
the equator the streamline starts, the larger the angle of rotation it makes
around the axis, the closer it gets to the central mass, and the nearer to  
the axis it asymptotically emerges. On the other hand, for streamlines 
beginning well above the equator, rotation is very nearly negligible and 
the path almost remains in a poloidal plane.   

%%%%%%%%%%%%%%%%%%%%%%%%%%%%%%%%%%%%
\begin{figure}
\psfig{figure=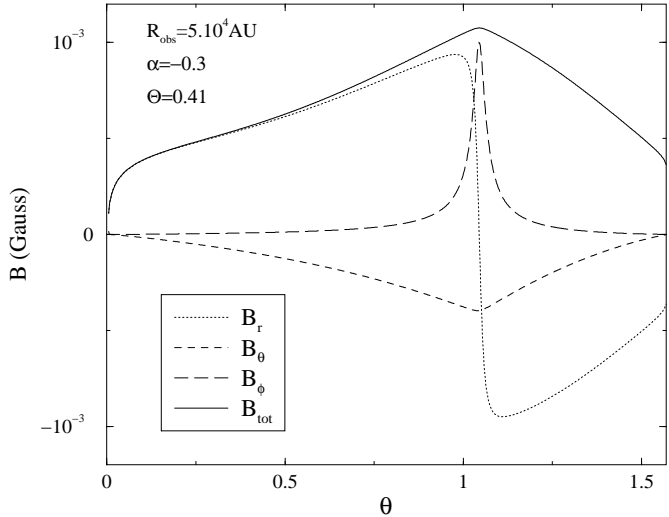,width=\linewidth}
\caption[ ]{
Virial-isothermal low mass solution: the angular dependence 
of magnetic field (in Gauss) at a distance of $5\times10^4$ AU.
Total magnetic field and $r$,$\theta$,$\phi$ components of $B$ are
represented. ($\alpha=-0.3$,$\Theta=0.41$,$M\approx1~M_\odot$)}
\label{figmAG}
\end{figure}
%%%%%%%%%%%%%%%%%%%%%%%%%%%%%%%%
Indirect measurements of the magnetic field in protostars are presently 
available (inferred from the radio flux from gyro-synchrotron emission by 
Ray et al. \cite{Ray}, and Hughes \cite{hughes}). Such measurements 
can provide a critical test of  models. In Fig.\ref{figmAG} the angular 
dependences of the components of the magnetic field are shown in Gauss at 
a distance of $5\times10^4$ AU for the same illustrative example of a low 
mass protostar. Magnetic fields of the order of a milligauss or less are 
obtained with a peak around 1 mGauss that coincides in angle with the 
radial turning point of the flow. As required by the quadrupolar geometry 
$B_\phi$ and $B_\theta$ vanish both on the equator and on the axis, while 
$B_r$ vanishes only on the axis and at the opening angle, where the largest 
field intensity exists.

%%%%%%%%%%%%%%%%%%%%%%%%%%%%%%%%%%%%
\begin{figure}
\psfig{figure=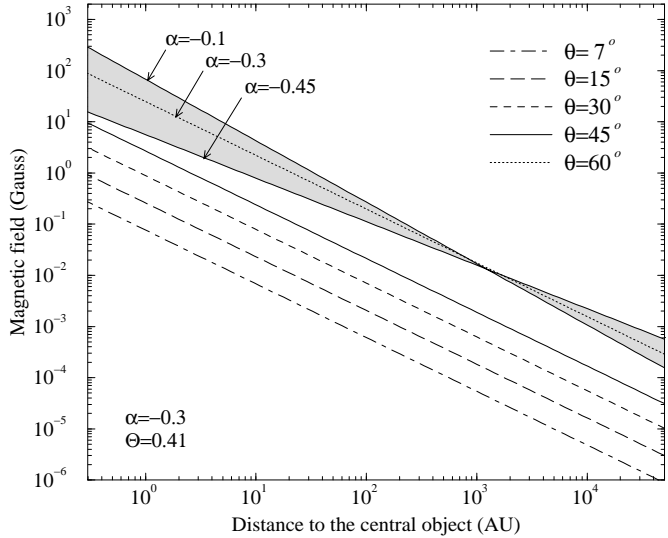,width=\linewidth}
\caption[ ]{
Virial-isothermal low mass solution ($\alpha=-0.3$,$\Theta=0.41$): 
magnetic field maximum (in Gauss) 
as a function of the distance to the central object (in AU)
for different angles ($\theta=7,15,30,45,60$).
The hashed region corresponds to the maximum magnetic field
for different self-similar index
($\alpha=-0.45,-0.3,-0.1$,$M\approx1~M_\odot$).}
\label{figdist}
\end{figure}
%%%%%%%%%%%%%%%%%%%%%%%%%%%%%%%%

One remarkable prediction of these models is that the magnetic field  
at a given radius varies dramatically with angle (and hence probably 
with inclination to the line of sight). The values range from 10 
microGauss to a milliGauss at $10^4$ AU, from $10^{-2}$ to a few Gauss 
at 20 to 40 AU in agreement with some observations (Ray et al. \cite{Ray}).
Peak field strengths may reach values as high as 1 to 100 Gauss at 1 AU 
(e.g. Hughes \cite{hughes}), as shown in Fig.\ref{figdist}. There is also a 
weaker dependence  on the self-similar index as is also shown in 
Fig.\ref{figdist}. These predicted magnetic field strengths are 
surprisingly large. {\em This is nevertheless consistent with existing 
(rare) observations and reinforce the idea that magnetic fields play 
a major role around young stellar objects} (Donati et al. \cite{donal97}, 
Guenther \& Emerson \cite{gue96}).

%%%%%%%%%%%%%%%%%%%%%%%%%%%%%%%%%%%%%%%%%%%%%%%%%%%%%%%%%%%%%
\subsection{Massive Protostars}
%%%%%%%%%%%%%%%%%%%%%%%%%%%%%%%%%%%%%%%%%%%%%%%%%%%%%%%%%%%%%

To adapt the model for high mass protostars,  we must find solutions 
with higher self-similar density (according to Eq.~\ref{high/low}) 
using the method outlined in Sect.~\ref{NUM} and adjust the other 
variables, mainly the magnetic field in fact, until the boundary 
conditions are satisfied. For radiative high 
mass protostars, we may speculate that we may not need a   
magnetic field as strong as in the low mass case to either launch
or collimate the outflows since strong pressure gradients are
produced by radiative heating (see also Shepherd \& Churchwell 
\cite{shch}, Churchwell \cite{church}). However the virial-isothermal model 
that we give below has a stronger magnetic field.

%%%%%%%%%%%%%%%%%%%%%%%%%%%%%%%%%%%%%%%%%%%%%%%%%%%%%%%%%%%%%
\subsubsection{The Virial-Isothermal Case \label{high:iso}}
%%%%%%%%%%%%%%%%%%%%%%%%%%%%%%%%%%%%%%%%%%%%%%%%%%%%%%%%%%%%%

%%%%%%%%%%%%%%%%%%%%%%%%%%%%%%%%%%%%
\begin{figure}
\psfig{figure=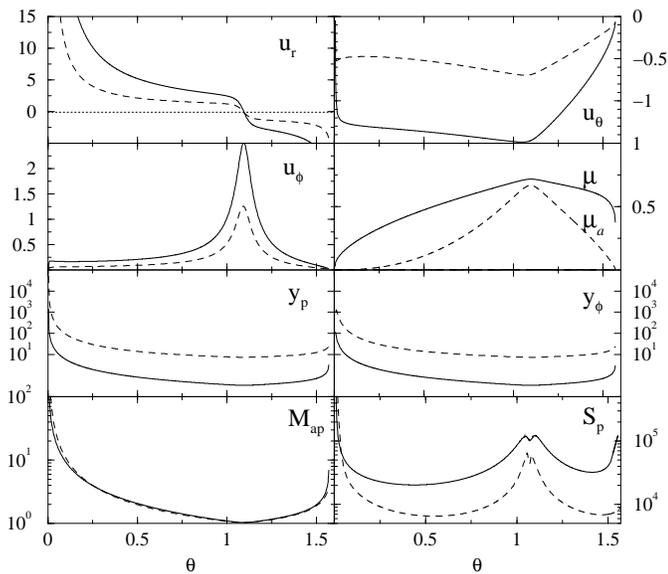,width=\linewidth}
\caption[ ]{
Massive protostar virial-isothermal solution:
Variations with angle of the self-similar variables (related to 
velocity ($u_r$, $u_\theta$, $u_\phi$), 
density ($\mu$ and $\mu_a$), magnetic field ($y_p$, $y_{\phi}$)
and Mach number ($M_{ap}$, $M_{a\phi}$)
 with $\alpha=-0.1$,$\Theta=0.1$ (solid lines).
The poloidal component $S_p$
of the Poynting flux is plotted in the lowest right panel.
A low mass solution ($M\approx1~M_\odot$) is plotted with dashed lines
over the high mass case ($M\approx30~M_\odot$).}
\label{figmassive}
\end{figure}
%%%%%%%%%%%%%%%%%%%%%%%%%%%%%%%%

Fig.\ref{figmassive} shows the self-similar variables for a 
virial-isothermal massive object. This solution is also completely  
super-Alfv\'enic, just as is the previous low mass example that is 
overplotted with dashed lines. Although the overall 
opening angle of the flow (as measured  by the angle at 
which the radial velocity changes sign) is the same in both cases, 
we observe that massive protostars produce faster (scaled) radial 
velocities throughout the positive range. There is also a more 
negative $u_\theta$ throughout this range which leads to an 
enhanced `collimation', corresponding to an increased axial density of 
material. In addition, the peak rotation at the `turning point' 
and the infall velocity just  
before this peak are both larger for the massive star. The smaller 
Alfv\'enic Mach numbers also show that the magnetic field is more 
important in this flow. This seems consistent with the ability to 
confine the {\it higher} physical temperature.

Our high mass virial-isothermal protostar yields larger scaled 
velocities than does the low mass object. This is mainly due to the larger 
luminosity in the high mass case. Sufficiently close to the axis both the 
low ($1~M_\odot$) and high ($30~M_\odot$) mass objects can produce 
$300~km~s^{-1}$. However as noted previously it is not clear that 
streamlines at such small angles are part of the present flow.

%%%%%%%%%%%%%%%%%%%%%%%%%%%%%%%%%%%%%%%%%%%%%%%%%%%%%%%%%%%%%
\subsubsection{The Radiative Case \label{high:rad}}
%%%%%%%%%%%%%%%%%%%%%%%%%%%%%%%%%%%%%%%%%%%%%%%%%%%%%%%%%%%%%

Massive molecular outflows present large bolometric luminosity, and
therefore radiation probably plays an important role in the
dynamics. Consequently radiative heating has to be 
investigated,  taking into account consistently the diffusion equation. 
The radiation field and the temperature are no longer constant with angle and 
 the radiation flux is not purely radial. The index $\alpha_f$ in
Eq.~\ref{selfrad} is chosen to be negative ($-0.1$ as in FH1) and so  
simulates a slight radiation loss. The numerical method is similar to 
that used above. The self-similar radiation flux $\vec f(\theta)$  is 
fixed at $\theta_{min}$ to be zero , and the non-zero value at 
$\theta_{max}$ measures the radiation `loss' from the self-similar region. 
%%%%%%%%%%%%%%%%%%%%%%%%%%%%%%%%%%%%
\begin{figure}
\psfig{figure=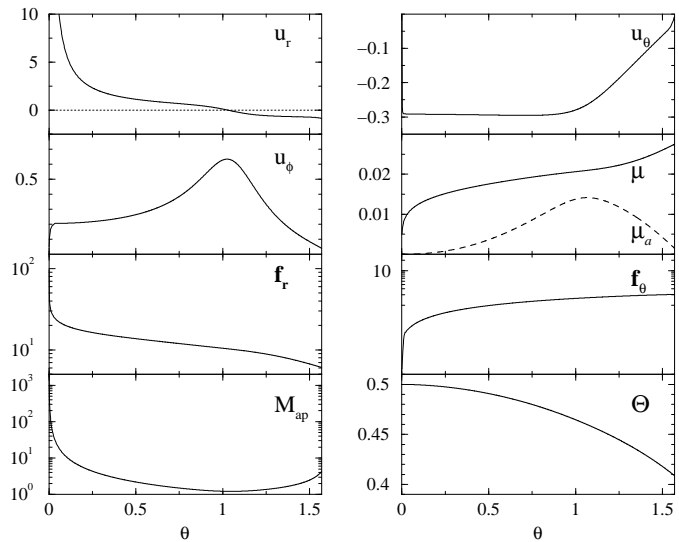,width=\linewidth}
\caption[ ]{
Massive protostar radiative solution:
Variations with angle of the self-similar variables related to 
velocity ($u_r$, $u_\theta$, $u_\phi$), 
density ($\mu$ and $\mu_a$), 
radiation field ($f_r$, $f_\theta$), Mach number ($M_{ap}$) and 
temperature ($\Theta$) for $\alpha=-0.2$.
$M\approx 5~M_\odot$.}
\label{figradSOL}
\end{figure}
%%%%%%%%%%%%%%%%%%%%%%%%%%%%%%%%
 
Fig.\ref{figradSOL} shows a solution for the radiative case
where $\alpha=-0.2$, $a=0$ and $b=2$. The temperature $\Theta$
is at its maximum (0.5) in the jet region and decreases towards the equator.
The components of the radiation flux are much smaller than in FH1.  Thus 
{\em the required heating is less important than in FH1}.

Fig.\ref{figCOLOR} presents two illustrative radiative solutions for low 
($M\approx 0.3~M_\odot$) and high mass ($M\approx 7~M_\odot$) cases. 
Streamlines in the poloidal plane are overplotted on contours of the hydrogen 
number density in the upper part of the figure, while the intensity of the 
radiation field is represented at the bottom. Fig.\ref{figCOLOR} clearly 
shows the saddle type singularity of the central point. We see that a  
Keplerian disc naturally appears as part of the global solution as does 
the collimated outflow.

%%%%%%%%%%%%%%%%%%%%%%%%%%%%%%%%%%%%
%\begin{figure*}
%FOR COLOR USE THE FOLLOWING
%\psfig{figure=RADtgif2.eps,width=\linewidth}
%\psfig{figure=RADlowQUAL.eps,width=\linewidth}
%\caption[ ]{
[[Caption of the figure \label{figCOLOR} 
that is given in gif format because of its size]
Streamlines and density contours of the hydrogen number density 
(in logarithmic scale) for low and high mass radiative solutions 
with $\alpha=-0.2$. The density levels shown set between $10^3$ 
to $10^5$ cm$^{-3}$ for the low mass case, and between $10^5$ to 
$10^7$ cm$^{-3}$in the other case. Length scales are in AU.
The lower panel presents the total intensity of the radiation field
(on a logarithmic scale).]
%}
%\label{figCOLOR}
%\end{figure*}
%%%%%%%%%%%%%%%%%%%%%%%%%%%%%%%%

The radiation field is smaller in the massive case in 
Fig.\ref{figCOLOR}.  This is remarkable since one should
expect significant radiative heating surrounding massive
protostars. However, we have chosen to have the same components 
of the self-similar velocity for both cases, as well as the 
same self-similar temperature and radiation flux as input 
parameters. Therefore Fig.\ref{figCOLOR} shows that  the 
necessary radiation field has to be larger in low mass
outflows in order to get the same scaled velocity as in the massive case.
Particularly in the latter case, the global shape of the radiation field 
is rather similar to the solution obtained by Madej et al. (\cite{MLH}),
where an anisotropic radiation field was produced by scattering in thick 
accretion discs. It was shown in their article that collimation was 
clearly displayed by the radiation field in the polar region. We also
find, in the massive case, that the radiation field increases away from the 
equator. On the other hand the low mass case almost shows an almost spherical 
symmetry, except on the axis.

\begin{figure}
%\picplace{6cm}
\psfig{figure=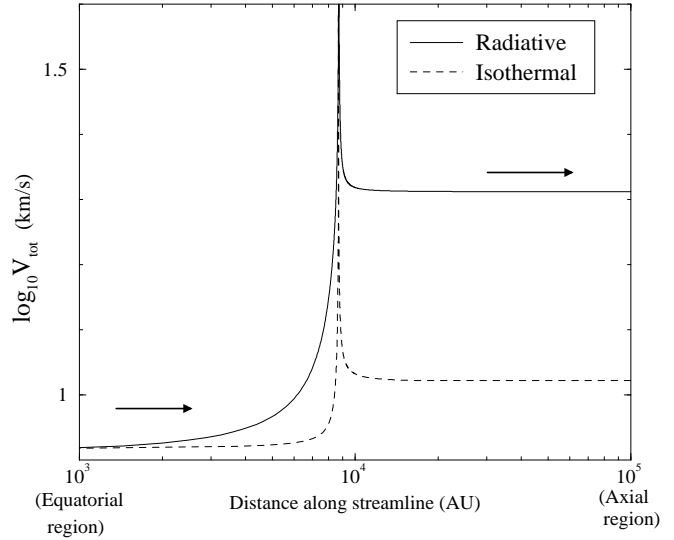,width=\linewidth}
\caption[ ]{
Variations of total velocity with integrated distance along a streamline
for radiative (solid) and virial-isothermal (dashed) cases ($\alpha=-0.2$,
$\Theta_{axis}=0.5$). The streamline are integrated from a radius of about 
$10^4$ AU in the equatorial region out to $10^6$ AU in the axial region, 
at a final angle of 0.1. The central object mass is $5~M_\odot$. 
Arrows denote the direction taken by the flow along streamline.
}
\label{figradcomp}
\end{figure}

Radiative heating increases the pressure gradient pushing outwards and
therefore peak and asymptotic velocities along streamlines are larger as 
represented in Fig.\ref{figradcomp}. This figure shows variation of 
total velocity with integrated distance along a streamline for both 
radiative and virial-isothermal cases for the same input parameters. 
We begin the integration in the equatorial region at about $10^4$ AU 
from the protostar and integrate out to $10^6$ AU in the axial region, 
where we attain  a final angle of 0.1 radian (see FH1 for comparison).
 Velocity is maximum at the turning point ($u_r=0$), 
which is the closest point to the source, and is 
almost completely toroidal there. The inclusion of radiative 
heating provides a net acceleration of material which clearly increases 
the asymptotic axial velocity.
  Moreover the asymmetry in this graph indicates that the angular 
variation dominates the radial decrease as dictated by the self-similar forms. 
In fact this asymmetry is the clearest indication of the production of 
high velocity outflow from the relatively slow (but massive) infall.
%%%%%%%%%%%%%%%%%%%%%%%%%%%%%%%%%%%%
%%%%%%%%%%%%%%%%%%%%%%%%%%%%%%%%

Global differences in the solutions between virial isothermal and radiative 
cases remain as described in FH1. The temperature is maximum in the jet 
region and decreases towards the equator. The radial component of the 
radiation flux decreases monotically in the same direction and its 
$\theta$-component increases, being always positive.

\begin{figure}
%\picplace{6cm}
\psfig{figure=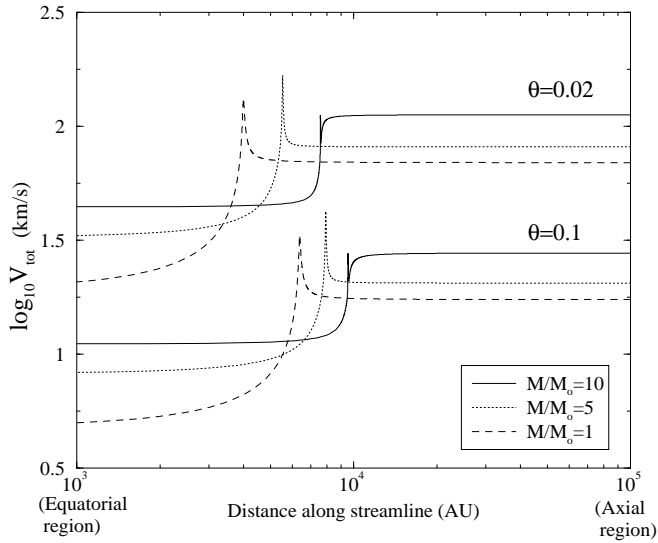,width=\linewidth}
\caption[ ]{
Variations of total velocity with integrated distance along a streamline
for protostars of different masses. Solid lines represent a central object 
of 10 solar masses while dotted and dashed respectively stands for 5 and 1 
$M_\odot$. Two sets of solutions are represented for two streamlines 
at different polar angles ($\theta=0.02;0.1$).
}
\label{figradALL}
\end{figure}

In order to show the influence of the mass on solutions, we present variations 
of the total velocity with the integrated distance along streamlines as 
defined in the previous plot for different masses in Fig.\ref{figradALL}.
For all the curves, the input parameters are kept the same  ($\alpha=-0.2$,
$\Theta=0.5$) except for the mass of the central object and therefore 
the self-similar density $\mu$. Even the initial components of velocity and 
magnetic field remain the same. Only the starting distance of integration 
from the central object varies in order to differentiate the various curves.
The larger luminosities for massive objects explain the larger velocities. 
In the solutions shown in Fig.\ref{figradALL}, the total velocity is 
multiplied by approximately a factor 5 as material flows from the 
equatorial region to the axial region. The efficiency of 
heating as a driving mechanism grows with the input values of the 
temperature  $\Theta$ and of the components of the self-similar 
radiation flux  $\vec f$. Indeed, a factor of 10 or more can be obtained 
just by increasing these parameters. Moreover, the figure shows that the 
highest velocity components ($\approx~100~km~s^{-1}$) are most tightly 
collimated along the axis ($\theta=0.02$). {\em Thus the most massive 
protostars produce the fastest flows with maximum radial velocities 
in the axial region}. 

%%%%%%%%%%%%%%%%%%%%%%%%%%%%%%%%%%%%
%%%%%%%%%%%%%%%%%%%%%%%%%%%%%%%%
We conclude that {\em radiative heating combined with Poynting flux 
driving provides an efficient mechanism for producing high velocity 
outflows} when the opacity is dominated by dust.

%%%%%%%%%%%%%%%%%%%%%%%%%%%%%%%%%%%%%%%%%%%%%%%%%%%%%%%%%%%%%
\subsection{Parameter Study}
%%%%%%%%%%%%%%%%%%%%%%%%%%%%%%%%%%%%%%%%%%%%%%%%%%%%%%%%%%%%%

%%%%%%%%%%%%%%%%%%%%%%%%%%%%%%%%%%%%
%\begin{figure}
%\picplace{6cm}
%\psfig{figure=graphUtpmax2.eps,width=\linewidth}
%\caption[ ]{
[[Caption of the figure \label{figmonte}
that is given in gif format because of its size]
Monte Carlo shooting:
Plots of maxima of $u_\phi$ at the turning point ($u_r=0$)
with respect to the maxima of $u_\theta$ for different values
of the temperature parameter ($\Theta$=0.01;0.16;0.26;0.36).
($\alpha=-0.2$)]
%}
%\label{figmonte}
%\end{figure}
%%%%%%%%%%%%%%%%%%%%%%%%%%%%%%%%

We use a Monte Carlo exploration of our parameter space in order to study 
general trends in the model. The self-similar index and the temperature 
are fixed and the six input parameters (velocity and magnetic field scaled 
components and the density) are randomly chosen. The conditions presented in
Sect.~\ref{NUM} are required for any solution to be considered `good'. One 
should note that there exists a broad range of solutions that possess 
favorable characteristics. Only the most significant results are reported 
here.

We plot the maxima of $u_\phi$ as functions  of maxima of $u_\theta$ for 
various values of the self-similar temperature ($\Theta$=0.01;0.16;0.26;0.36),
and for a given index ($\alpha=-0.2$). These maxima occur at 
the turning point where the radial velocity vanishes. 
Each point shown here represents a `good' solution. 
Rotation is directly measured by $u_\phi$, and the tendency for 
material to go towards the axis, i.e. the collimation, is related to 
$u_\theta$. We find, as a global trend, that
when $(u_\theta)_{max}$ decreases, $(u_\phi)_{max}$ remains almost constant
($(u_\phi)_{max} \approx 0.25 (u_\theta)_{max}+constant$)
until $(u_\theta)_{max}$ reaches a threshold value. Then $(u_\phi)_{max}$
decreases faster than $(u_\theta)_{max}$, which is what was anticipated 
analytically for super-Alfv\'enic flow when $\alpha=1/4$ (see appendix B). 
In addition, there 
is a limiting $(u_\phi)_{max}$ for each $(u_\theta)_{max}$ at a 
given temperature, below which solutions are not found.  
When $\Theta$ increases, which physically corresponds to larger pressure 
gradients, $(u_\phi)_{max}$ decreases while $(u_\theta)_{max}$ 
becomes less negative. If the magnitude of the latter quantity is taken as 
a measure of the collimation, we see that {\em rotation and 
collimation decrease together with increasing temperature and therefore 
probably with the bolometric luminosity of the central object}. 

It is found that solutions are less influenced by density (and hence mass) 
than by temperature. Nevertheless the distribution of opening angle is not 
found to change its form with temperature.
In particular the most probable value is always found essentially 
at the angle shown in the two sample solutions. The variation is such that 
the larger opening angles are found at the lower left of the band for each 
temperature, that is with maximal poloidal collimation flow and  the 
smallest possible maximum rotation.  

It is also found that {\em larger opening angles are associated with  
smaller magnetic fields}. The corresponding results are not reported here, but 
in fact the variation of the opening angle is rather large for a relatively 
small magnetic field variation (a reduction of 25$\%$ in the magnetic field
magnitude can widen the opening angle from $35^\circ$ to $60^\circ$).
If the magnetic field varies secularly as the evolution proceeds, then 
a sequence of our models could be regarded as a series of `snapshots' of the 
protostellar evolution. This evolutionary sequence would show that the 
opening angle increases with time, as the magnetic field becomes less 
important.  This would be consistent with the notion that the youngest 
outflows are generally the most collimated. Indeed, this has been 
observationally verified by  Velusamy and Langer (\cite{VeluLan}), who 
provide evidence that outflows widen in time.  Moreover when the magnetic 
field reaches a lower threshold in our model, the solution becomes a pure 
outflow. Therefore, as obtained with our model, if the opening angle 
broadens sufficiently, it may ultimately cut off the accretion supplying 
new material for the outflow. 

%%%%%%%%%%%%%%%%%%%%%%%%%%%%%%%%%%%%
\begin{figure*}
%\picplace{6cm}
\psfig{figure=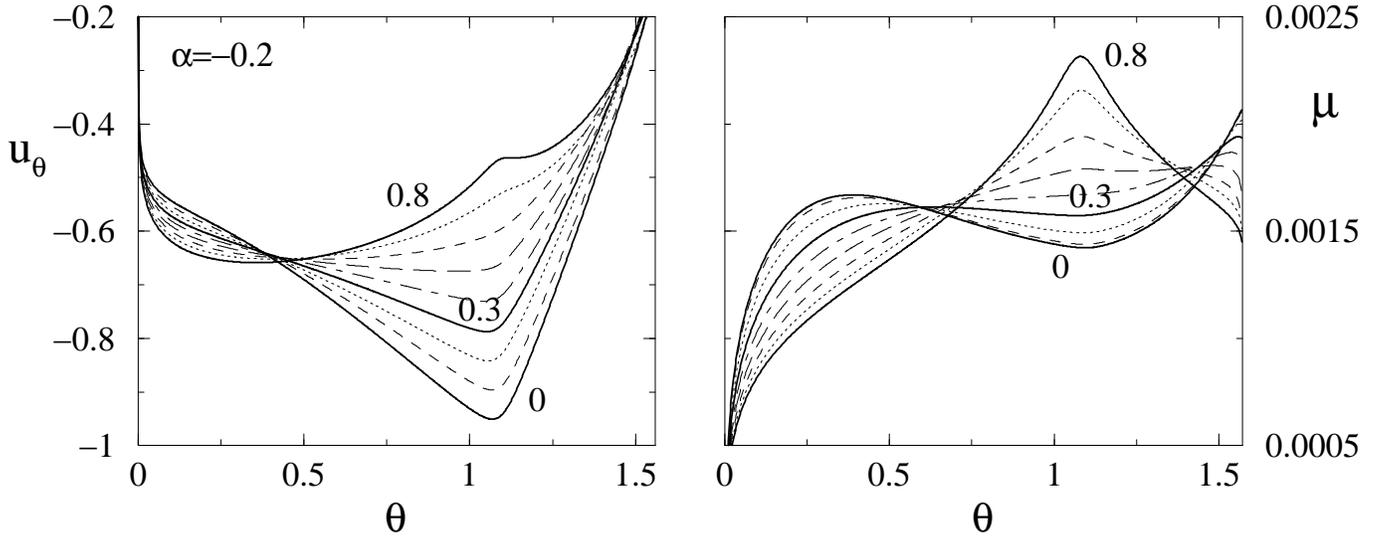,width=\linewidth}
\caption[ ]{
The angular dependence of $u_\theta$ (left panel) and of
density $\mu$ (right panel) for various self-similar 
temperature parameters (with $\alpha=-0.2$).
Solutions with $\Theta=0;0.3;0.8$ are represented with solid lines.
Two consecutive solutions are separated by a change of 0.1
in temperature.
}
\label{figuTHETA}
\end{figure*}
%%%%%%%%%%%%%%%%%%%%%%%%%%%%%%%%
The angular dependence of the self-similar variables are represented in
Fig.\ref{figuTHETA}. The only parameter that varies is the self-similar 
temperature. The figure (left panel) illustrates the fact that collimation 
decreases with temperature as already mentioned. Moreover variations of 
the self-similar temperature give rise to different density profiles in 
the infalling `disc' region, as seen on the right hand panel of 
Fig.\ref{figuTHETA}. The density decreases close to the equator for the 
largest values of the temperature with a maximum well away from the equator. 
Such solutions accrete so rapidly in the equatorial plane that they have a 
reduced density relative to their rotationally supported surroundings. 
These solutions resemble the special case $\alpha=1/4$ as discussed in 
appendix B. On the other hand the temperature has little effect on the 
density profile in the axial region, although it broadens the central 
`funnel' substantially.

We note that solutions for cold flows ($\Theta=0$) have also been obtained.
In such a case, the circulation is powered only by Poynting flux since our 
Bernoulli integral (Eq.~\ref{Bernoulli}) applies. We do not expect asymmetric 
velocities at the two infinities on the stream lines. For the same set of 
parameters, the total velocity in the `interacting' region is smaller for 
cold flows and the fast rotating region (where $u_\phi$ is maximum) is 
narrower relative to finite pressure flows. 
Moreover the Alfv\'enic Mach numbers almost reach unity at the 
peak rotation point, showing that the flow is only slightly super-Alfv\'enic 
there, and that in fact the two terms in Eq.~\ref{Bernoulli} are probably 
comparable.

%%%%%%%%%%%%%%%%%%%%%%%%%%%%%%%%%%%%
\begin{figure}
%\picplace{6cm}
\psfig{figure=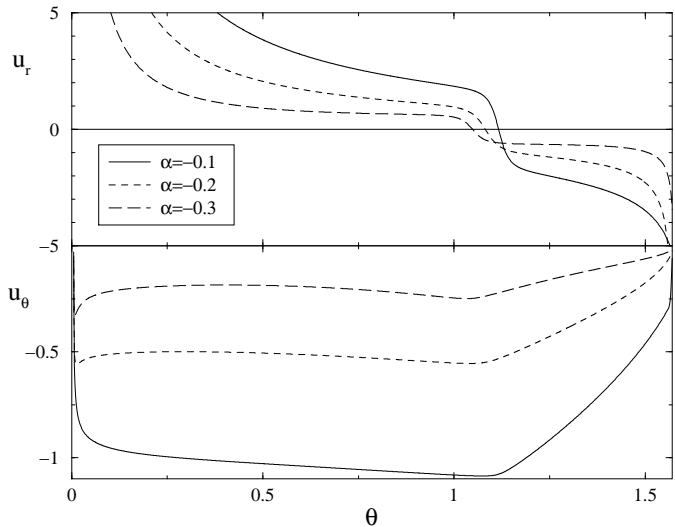,width=\linewidth}
\caption[ ]{
Dependences on the self-similar index
$\alpha=-0.1,-0.2,-0.3$ (with $\Theta=0.3$).
Radial (upper panel) and toroidal (lower panel)
components of the self-similar velocity are plotted
as functions of the angle. 
}
\label{figALPHA}
\end{figure}
%%%%%%%%%%%%%%%%%%%%%%%%%%%%%%%%
Dependences  of the self-similar index have also been investigated. 
The main results are illustrated by Fig.\ref{figALPHA} where the radial 
and longitudinal components of velocity are plotted for three values 
of the self-similar index but with the same self-similar temperature. 
For more negative self-similar index one gets dramatically smaller 
$\theta$- and $\phi$- velocities (recall that $\alpha=-1/2$ yields 
pure radial accretion) and a smaller change in the radial velocity.

%%%%%%%%%%%%%%%%%%%%%%%%%%%%%%%%%%%%%%%%%%%%%%%%%%%%%%%%%%%%%
\section{Observational Consequences of the Model: Synthetic 
$^{13}CO$ Spectral Lines and Maps.}
%%%%%%%%%%%%%%%%%%%%%%%%%%%%%%%%%%%%%%%%%%%%%%%%%%%%%%%%%%%%%

In this Section, we compute $^{13}CO$ (J=$1\rightarrow 0$) emission lines for
several of the outflow solutions discussed in Sect.~\ref{NUM}. Specifically,
we provide examples of spectra, channel maps, position-velocity diagrams, 
and intensity-velocity diagrams for one example of each of the following: 
the low mass solution of Sect.~\ref{low:iso}, the radiative high mass 
solution of Sect.~\ref{high:rad}, and the virial-isothermal high mass 
solution of Sect.~\ref{high:iso}.  
The importance of this analysis is that it allows us to directly 
compare the observable features of our models with real observational 
data.  However, we cannot fully explore the observational consequences 
of our models in the present paper.  Our purpose in this Section will 
be mainly to outline the qualitative features of our models, leaving a 
complete exploration of the parameter space for the second paper in our 
series. 

The method that we follow is essentially the same as that outlined in 
FH2, except that the code used there has been substantially improved 
in a number of respects. Firstly, we are now able to generate results 
with much higher spectral and spatial resolution. Secondly, although 
we compute spectra on a grid of pencil beams through the outflow source, 
we convolve our spectra with a Gaussian telescope beam to more 
accurately simulate observational results.  FH2, on the other hand, 
did not perform this convolution, which resulted in many spuriously 
sharp features in their maps.  Thirdly, we do not embed our solutions 
in a background of molecular gas to simulate the molecular cloud, 
which presumably surrounds the outflow.  FH2 demonstrated that the 
presence of background material has no significant effect on the 
spectra or maps, except for in the lowest, and hence least interesting, 
velocity channels.  Since we are primarily interested in emission at 
relatively high velocities, corresponding to the outflow, emission or 
absorption by relatively slow moving background gas is of no consequence.  
As in FH2, the primary limitation of our analysis is that we assume 
local thermodynamic equilibrium.  Although this is probably not strictly 
true, we do expect the $^{13}CO$ level populations to be approximately 
in equilibrium with the local temperature provided that the optical 
depth is at least moderately high.  Therefore, we hope to capture at 
least the character of the emission, if not the precise intensity.

\subsection{Low Mass Solutions} \label{sec:low}

\begin{figure}
\psfig{file=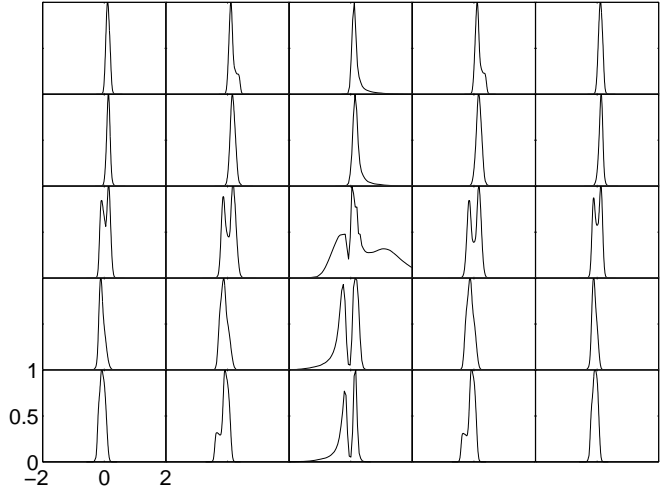,width=\linewidth}
\caption{Normalized spectra are shown for the low mass solution on a 
$5\times 5$ grid of map positions that are evenly spaced across the 
mapped region, which is $20\times 40~r_o$ in size. The abscissa is in 
units of $km~s^{-1}$. The source is located at the center.}
\label{fig:LOWspectra}
\end{figure}

We have assumed a mass of $1~M_\odot$ for the low mass solution, 
which implies a fiducial radius of $r_o=1400~AU$ using Eq.~\ref{ro}.  
We have also chosen a inclination angle of $45^\circ$ between the symmetry 
axis of the outflow and the plane of the sky.  With the mass and $r_o$ 
determined, the density, temperature, and velocity are determined at any 
position by the dimensionless parameters $\mu(\theta)$ and $\Theta(\theta)$, 
and the radial scalings given by Eqs.~\ref{self1}-\ref{selfrad}.  
Therefore, by constructing a line of sight through the cloud, we can compute 
the emission (expressed as a radiation temperature) as a function of velocity 
parallel to the line of sight; examples of these spectra are shown in 
Fig.\ref{fig:LOWspectra}.  We have computed spectra on a $61\times 120$ 
grid of map positions across the outflow.
%%%%%%%%%%%%%%%%%%%%%%%%%%%%%%%

\begin{figure}
\psfig{file=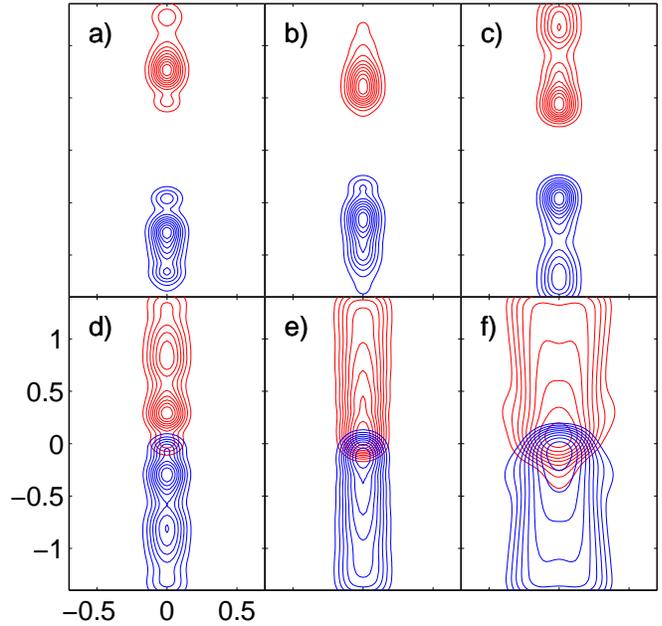,width=\linewidth}
\caption{Channel maps for the low mass solution at an inclination 
angle of $45^\circ$.  In units of $km~s^{-1}$, the velocity channels are 
as follows for panels a) through f): a) $5.5<|v|<6.5$, b) $4.5<|v|<5.5$, 
c) $3.5<|v|<4.5$, d) $2.5<|v|<3.5$, e) $1.5<|v|<2.5$, f) $0.5<|v|<1.5$.  
The upper part of the outflow is the red-shifted side, while the lower 
part is blue-shifted.  Map coordinates are in units of $10^{4}~AU$. }
\label{fig:LOWmap}
\end{figure}

\begin{figure}
\psfig{file=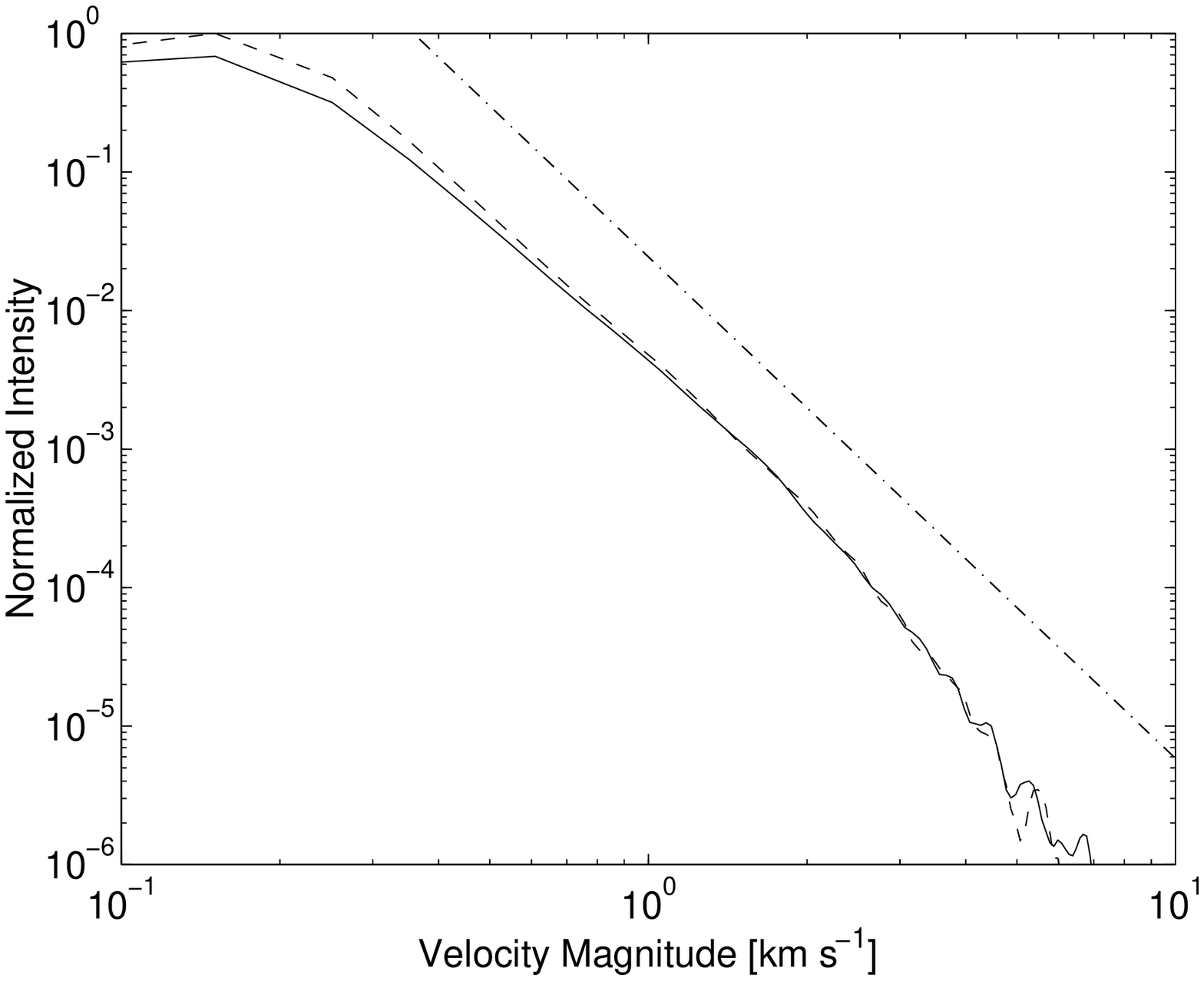,width=\linewidth}
\caption{Intensity-velocity diagram for the low mass solution.  
The dot-dashed line represents a power-law fit with an index of $-3.5$.}
\label{fig:LOWiv}
\end{figure}

\begin{figure}
\psfig{file=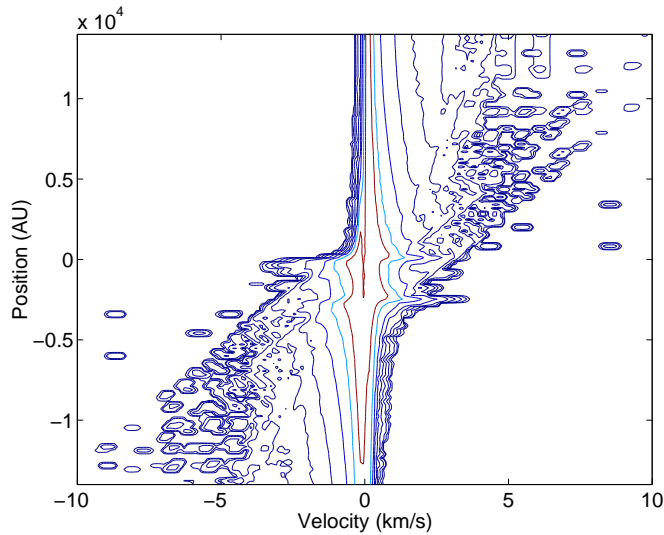,width=\linewidth}
\caption{Position-velocity diagram for the low mass solution.}
\label{fig:LOWpv}
\end{figure}
%%%%%%%%%%%%%%%%%%%%%%%%%%%%%%%%

We find that the spectra shown in Fig.\ref{fig:LOWspectra} vary substantially 
with map position across the outflow.  Spectra computed near the map 
projection of the outflow axis typically show relatively high velocity wings
that may extend up to several $km~s^{-1}$, but the emission generally becomes
too weak to show past $\approx 2~km~s^{-1}$.  We note that the strongest 
emission always occurs at low velocities, with relatively weak emission in 
the wings, as is the case for real outflows.  Spectra may contain either a 
single wing or both red and blue-shifted wings depending on whether a 
particular line of sight crosses only a single lobe of the outflow or both. 
There are also many cases where line wings are absent, but well-defined 
``shoulders'' are present on the low velocity dominated spectra.  

In the equatorial regions of the maps, double-peaked spectra are found with 
relatively low velocity peaks. These spectra are mainly due to the radial 
infall that dominates in the equatorial regions.  It is unlikely that the 
double-peaked profiles are due to rotation, since radial infall velocities 
exceed the rotational velocities at all angles $\theta$ 
except for very near the radial turning point of the outflow.
Furthermore, many of the spectra have a slight asymmetry in which 
the blue-shifted peak shows stronger emission.  This is suggestive of the
well-known outflow asymmetry, which is due to the more efficient 
self-absorption of blue-shifted emission by overlying material moving 
towards the observer (Shepherd \& Churchwell \cite{shch}, 
Gregersen et al. \cite{greg}, Zhou et al. \cite{zhou}). 
The effect is slight in this case since the low 
mass solution, with a $1~M_\odot$ central object, produces a peak optical 
depth of only $\approx 0.2$ along most lines of sight.  The massive solutions, 
discussed in Sect.~\ref{high:rad} and \ref{high:iso}, produces much 
higher optical depths and leads to a much more pronounced asymmetry.

Fig.\ref{fig:LOWmap} shows a set of channel maps for the low mass 
solution.  To make these synthetic maps, we have integrated the emission 
over $1~km~s^{-1}$ wide channels from $-6.5$ to $6.5~km~s^{-1}$.  We have 
deliberately excluded the emission with $|v|<1~km~s^{-1}$, since we find 
that the lowest velocity emission is essentially featureless.  We have 
convolved the maps with a Gaussian telescope beam with a FWHM of $800~AU$; 
for example, this corresponds to a $5~arcsec$ beam at a distance of $160~pc$.

The outflow is apparent at all velocities shown in our maps.  We obtain a 
relatively wide outflow ``cone'' at the lowest velocities ($0.5$ to 
$1.5~km~s^{-1}$) and very well collimated jet-like emission apparent at 
velocities between $1.5$ and $3.5~km~s^{-1}$. At higher velocities, the 
jet-like feature breaks up into outflow ``spots'' that move away from the 
central object as the velocity increases.  {\em The most important feature 
of these channel maps is that the opening angle of the outflow gradually 
decreases as the magnitude of the velocity increases.} As discussed in FH2, 
this effect is entirely due to the angular velocity sorting of the outflow 
solutions.  Our models {\em always} have the property that the outflow 
velocity increases towards the axis of symmetry.  {\em Thus, our models 
naturally produce outflows in which most of the material is poorly 
collimated and moves at relatively low velocities, but the fastest jet-like 
components are very well collimated towards the axis of symmetry.} This 
property has, in fact, been observed in a number of outflow sources 
(Bachiller et al. \cite{bachiller}, Guilloteau et al. \cite{Guilloteau}, 
Gueth et al \cite{Gueth}).

Several authors  have noted that molecular outflows seem to be characterized 
by a power-law dependence of total integrated intensity as a function of 
velocity (Rodriguez et al. \cite{Rodriguez}, Masson \& Chernin \cite{Masson}, 
Chandler et al. \cite{Chandler}). In Fig.\ref{fig:LOWiv}, we have 
summed all of the spectra in order to show how the intensity is distributed 
with velocity.  The solid line in the figure represents the blue-shifted 
emission, while the dashed line represents the red-shifted emission. We note 
that there are no important differences due to the somewhat low optical 
depth of our solution.  We find that the intensity $I$ is well fit by a 
power law $I\propto v^{-3.5}$ over all velocities greater than approximately 
$0.2~km~s^{-1}$.  This power-law behavior has, in fact, been found for 
several real outflows , and the index is in reasonable agreement with the 
available data (Cabrit et al. \cite{cab}, and Richer et al. \cite{PPIV}). 
It remains to be seen, however, how sensitive the power-law index is to the 
parameters of our model.  This important issue will be resolved in the 
second paper in this series, where we more completely explore the line 
emission of our models.

An important feature of Fig.\ref{fig:LOWiv} is that the emission falls 
{\em very} rapidly with increasing velocity.  For example, the intensity 
at even a rather modest velocity of $3~km~s^{-1}$ is only $0.01\%$ of the 
peak intensity.  At some velocity, the emission will fall below the noise 
threshold that is always present in real observations.  Past this velocity, 
the emission shown in Fig.\ref{fig:LOWmap} would be undetectable.  

In Fig.\ref{fig:LOWpv}, we show a velocity-position diagram for a cut 
along the outflow axis.  We have actually computed spectra along several 
cuts parallel to the outflow axis and convolved them with a Gaussian beam 
to produce a single velocity-position diagram. We find that there is 
substantial emission at low velocities along the entire cut.  The highest 
velocities are found near the map centre, as must be the case since 
$v\propto r^{-1/2}$ by Eq.~\ref{self1}. We note that the emission 
contours in this figure are logarithmically spaced; thus, we find that 
the most intense emission occurs at low velocities at all map positions.  

The outflow is apparent in Fig.\ref{fig:LOWpv} by the emission extending 
out to approximately $\pm 8~km~s^{-1}$ on either side of the outflow.  
Moreover, the general appearance of the position-velocity diagram is 
suggestive of a globally accelerated flow since higher velocities
generally appear at greater distances from the central source.  How can 
this be reconciled with the fact that velocity decreases with radius in 
our model?  It was shown in FH2 that the apparent acceleration is due to 
the unique velocity sorting that is predicted by our model.  The radial 
velocity always increases with decreasing polar angle near the outflow 
axis. Our cut crosses progressively smaller angles with increasing 
distance from the central object, and therefore encounters higher 
velocity material. 

The ragged appearance of the highest velocity contours is entirely due 
to the limitations of our numerical procedures at high velocities. In 
our model, the highest velocity components are extremely localized near   
the central star and the outflow axis.  Thus, whether or not a pencil beam
passes through such a component is a matter of chance when the spatial   
extent falls below the grid spacing.  This apparently happens when 
$|v|\gtrsim 5~km~s^{-1}$.

\subsection{Massive Solutions}
\subsubsection{The Radiative Case} \label{sec:rad}
%%%%%%%%%%%%%%%%%%%%%%%%%%%%%%%     
\begin{figure} 
\psfig{file=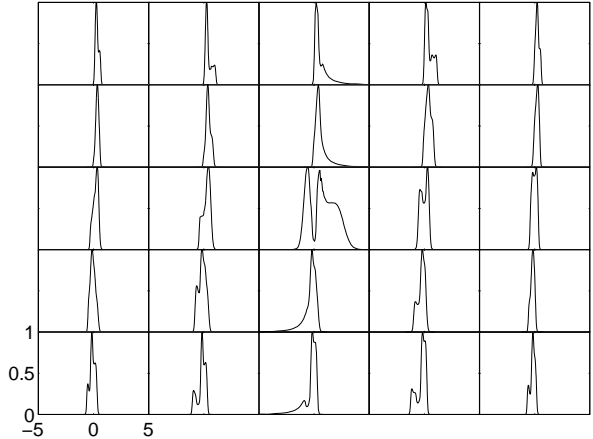,width=\linewidth}
\caption{Normalized spectra are shown for the radiative solution on
a $5\times 5$ grid of map positions that are evenly spaced across the 
mapped region, which is $10\times 20~r_o$ in size.
The abscissa is in units of $km~s^{-1}$.}
\label{fig:RADspectra}
\end{figure}

\begin{figure}
\psfig{file=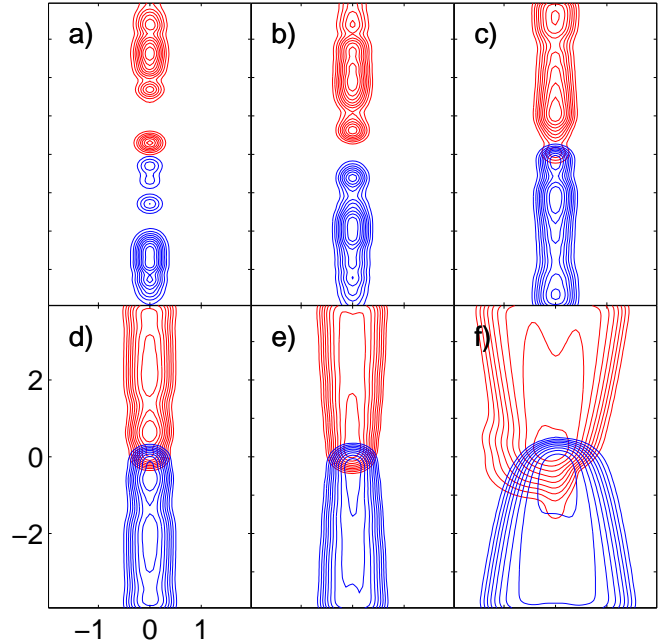,width=\linewidth}
\caption{Channel maps for the radiative solution at an inclination 
angle of $45^\circ$. In units of $km~s^{-1}$, the velocity channels are
as follows for panels a) through f): a) $11<|v|<13$, b) $9<|v|<11$, c) 
$7<|v|<9$, d) $5<|v|<7$, e) $3<|v|<5$, f) $1<|v|<3$.  The upper part of 
the outflow is the red-shifted side, while the lower part is blue-shifted.
Map coordinates are in units of $10^{4}~AU$. }
\label{fig:RADmap}    
\end{figure}

\begin{figure}
\psfig{file=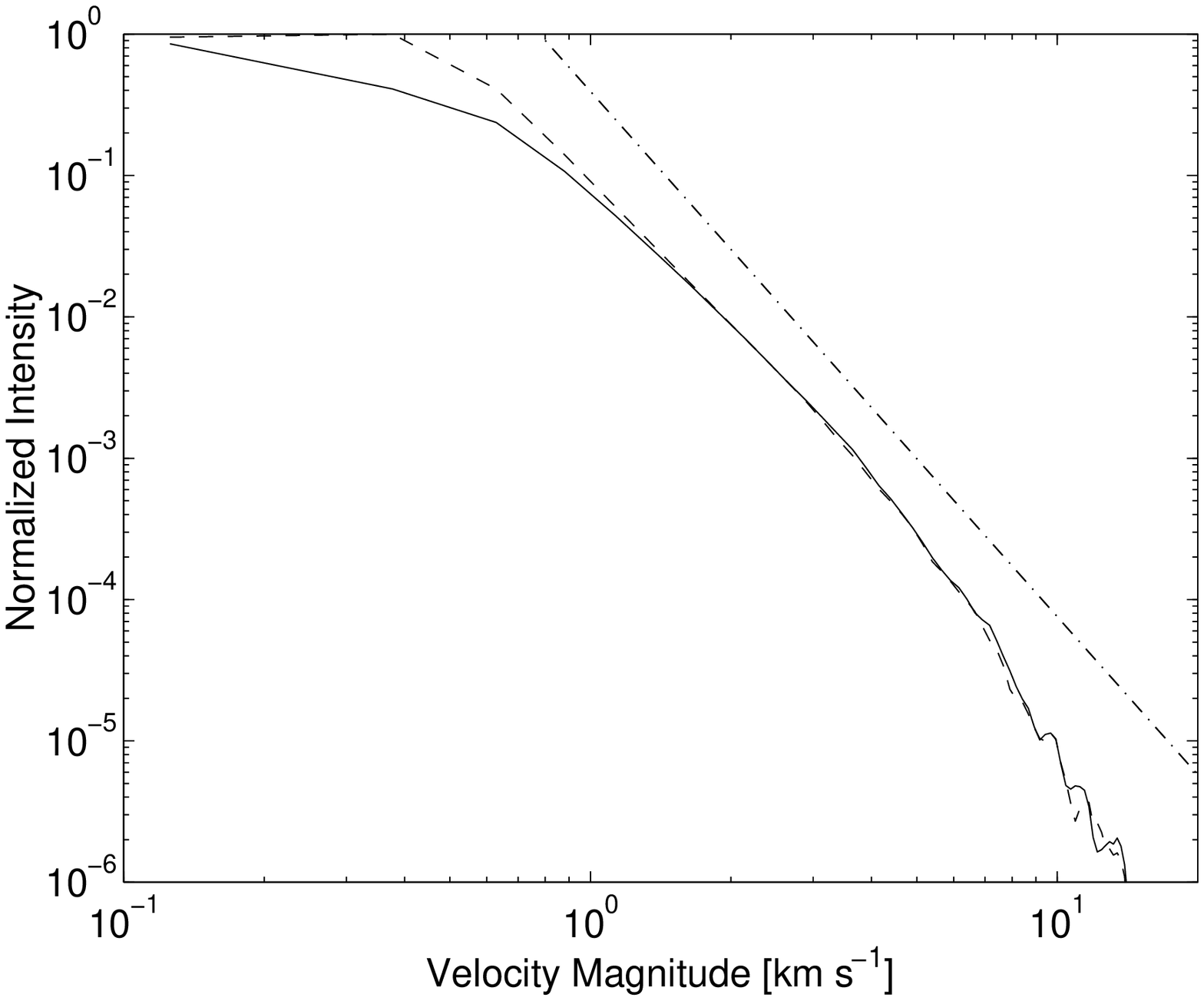,width=\linewidth}
\caption{Intensity-velocity diagram for the radiative solution.  
The dot-dashed line represents a power-law fit with an index of $-3.6$.}
\label{fig:RADiv}     
\end{figure}

\begin{figure}
\psfig{file=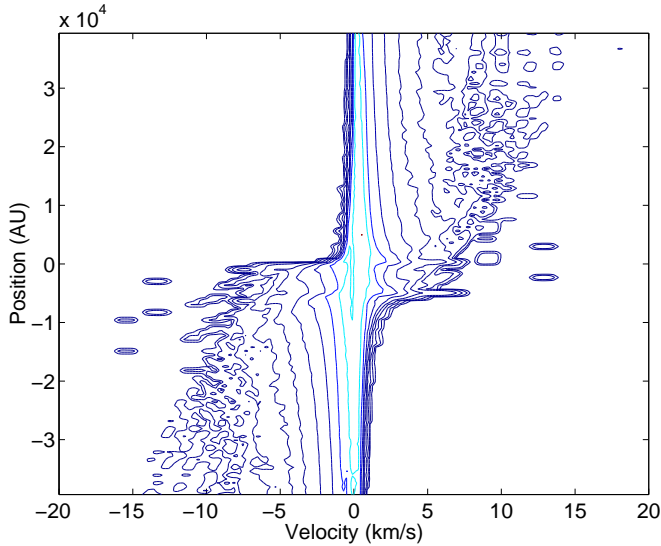,width=\linewidth}
\caption{Position-velocity diagram for the radiative solution.}
\label{fig:RADpv}     
\end{figure}
%%%%%%%%%%%%%%%%%%%%%%%%%%%%%%%%

It is useful to compare the $^{13}CO$ emission of the radiative solution 
discussed in Sect.~\ref{high:rad}, with the low mass solution discussed 
above.
We have assumed a mass of $10~M_\odot$, which implies a fiducial radius of 
$r_o=7900~AU$ according to Eq.~\ref{ro}. We find, heuristically, that 
the peak optical depth in the line centre decreases with increasing central 
mass.  Since the mass is a free parameter, we have tuned it to achieve 
realistic optical depths of order unity. The mass is larger than for the 
low mass solution discussed above, but the effects of radiative heating 
are more likely to become important in outflows surrounding more massive 
protostars, in any case.  The only other parameter that we have changed is 
the FWHM of the Gaussian beam that we convolve with our solution; here, we 
can use a slightly larger and more realistic 10 arcsec beam without smearing 
out the details of the solution. As in Sect.~\ref{sec:low}, we have 
computed spectra on a $61\times 120$ grid of map positions assumed a 
$45^\circ$ inclination angle.

Qualitatively, we find a great deal of similarity with the low mass solution.
The line profiles shown in Fig.\ref{fig:RADspectra} have similar
structure, and the outflow shown in Fig.\ref{fig:RADmap} is collimated to 
a similar degree.  The main difference is the velocity of the flow.  In the 
radiative case, we find that there is significant emission out to at least 
$10~km~s^{-1}$, and we find some low intensity emission out to 
$15~km~s^{-1}$; this is most clearly illustrated by the position velocity 
diagram shown in Fig.\ref{fig:RADpv}. Examining the intensity-velocity 
diagram shown in Fig.\ref{fig:RADiv}, we find that the intensity 
$I\propto v^{-3.6}$ for velocities greater than about $0.5~km~s^{-1}$.
It is striking that nearly the same power law index should be obtained for 
both the low mass and radiative solutions.  Admittedly, we do not fully 
understand the reasons for this similarity in behavior at present.  We also 
doubt that this sort of behavior is a generic feature of our models.
Nevertheless, we are encouraged that at least some of our models are 
capable of reproducing such realistic observed properties.  

\subsubsection{The Virial-Isothermal Case} \label{sec:mass}

%%%%%%%%%%%%%%%%%%%%%%%%%%%%%%%     
\begin{figure} 
\psfig{file=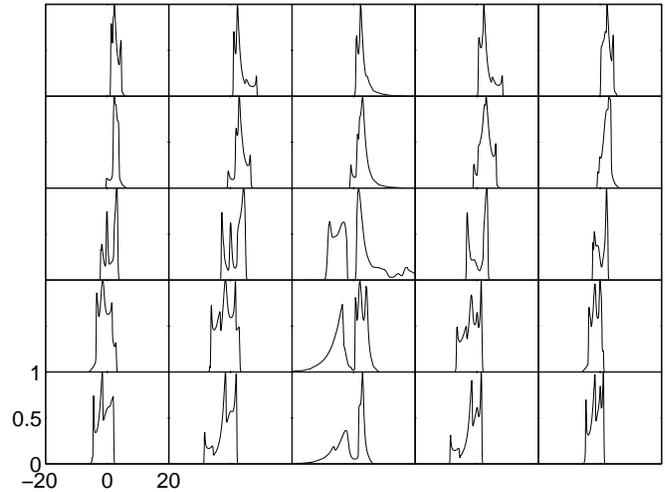,width=\linewidth}
\caption{Normalized spectra are shown for the massive solution on a 
$5\times 5$ grid of map positions that are evenly spaced across the 
mapped region, which is $2.5\times 5~r_o$ in size.
The abscissa is in units of $km~s^{-1}$.}
\label{fig:MASSspectra}
\end{figure}

\begin{figure}
\psfig{file=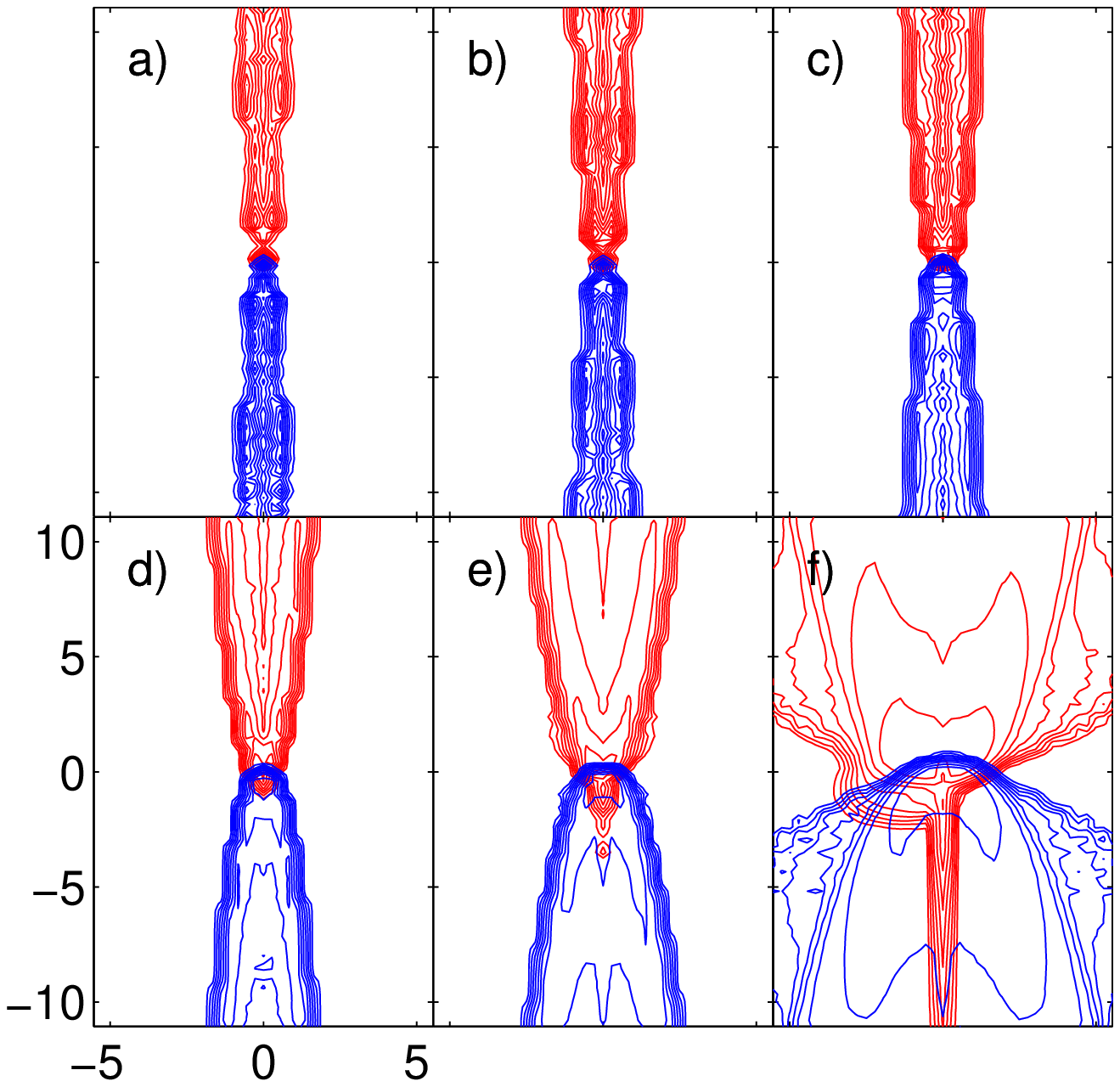,width=\linewidth}
\caption{Channel maps for the massive solution at an inclination 
angle of $45^\circ$. In units of $km~s^{-1}$, the velocity channels are
as follows for panels a) through f): a) $30<|v|<35$, b) $25<|v|<30$, 
c) $20<|v|<25$, d) $15<|v|<20$, e) $10<|v|<15$, f) $5<|v|<10$.  The 
upper part of the outflow is the red-shifted side, while the lower part 
is blue-shifted.  Map coordinates are in units of $10^{4}~AU$. }
\label{fig:MASSmap}    
\end{figure}

\begin{figure}
\psfig{file=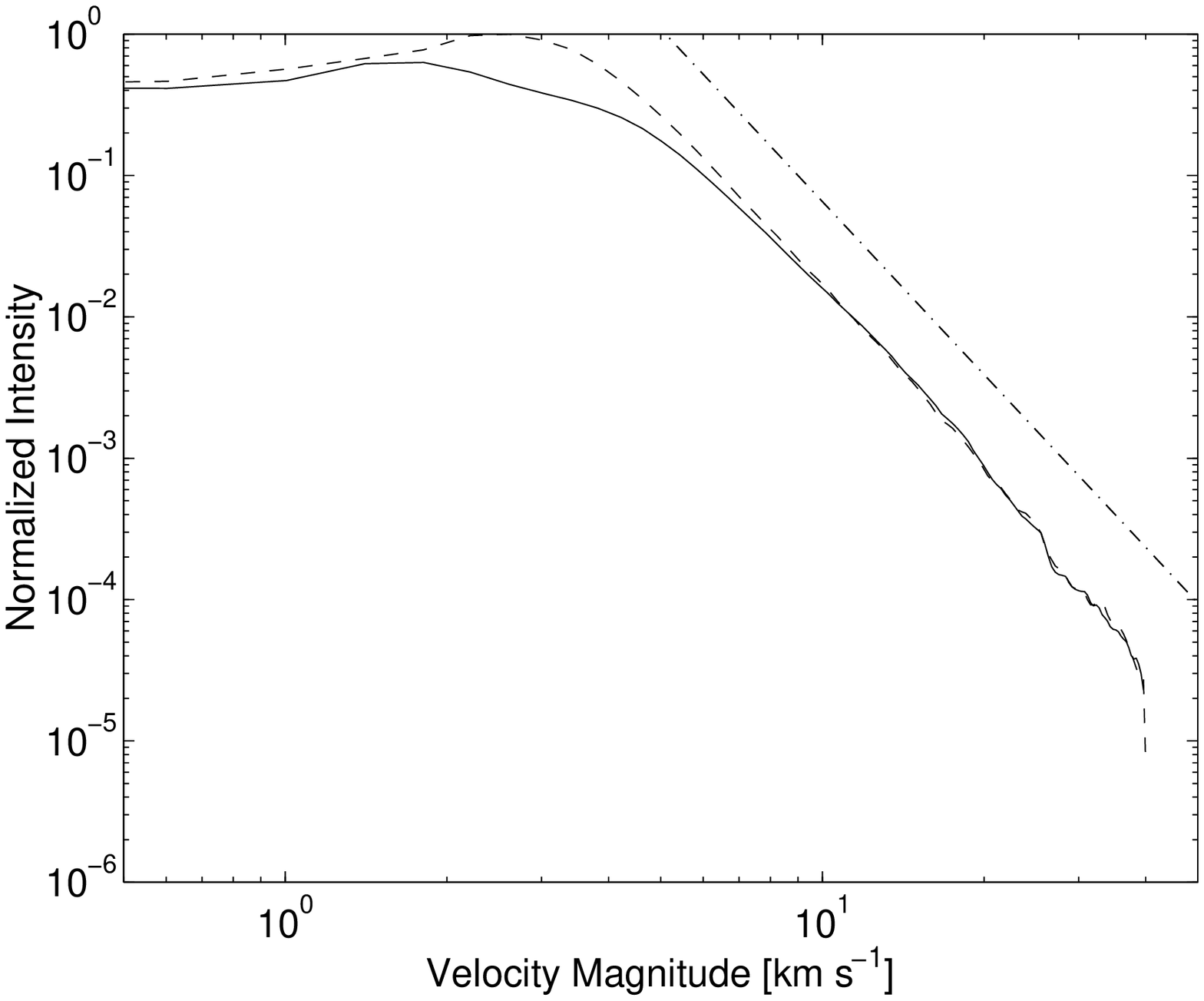,width=\linewidth}
\caption{Intensity-velocity diagram for the massive solution.  
The dot-dashed line represents a power-law fit with an index of $-4.1$.}
\label{fig:MASSiv}     
\end{figure}

\begin{figure}
\psfig{file=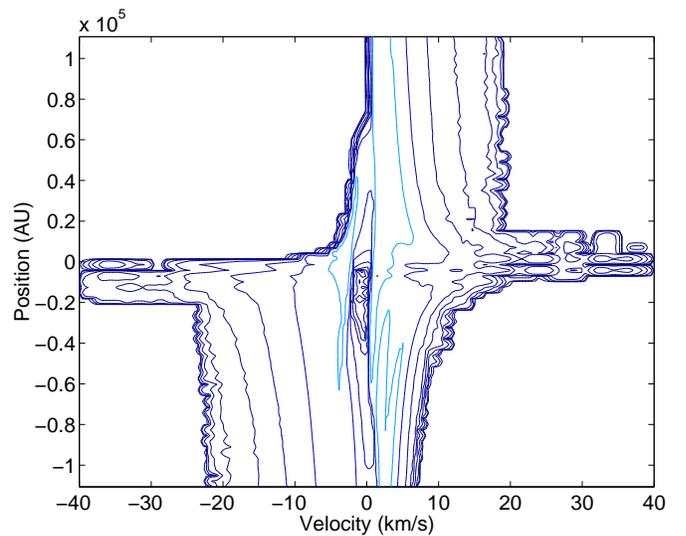,width=\linewidth}
\caption{Position-velocity diagram for the massive solution.}
\label{fig:MASSpv}     
\end{figure}
%%%%%%%%%%%%%%%%%%%%%%%%%%%%%%%%

We have also computed the $^{13}CO$ emission of the high mass solution 
discussed previously in Sect.~\ref{high:iso}. We find that the optical depth 
is extremely high unless the mass of the central protostar is chosen to be 
rather large; therefore, we have used a mass of $100~M_\odot$ and a 
corresponding fiducial radius of $r_o=44000~AU$ to compute the spectra. 
Even using such a large central mass, we find that the peak optical depth 
in the line centre is very high, with values exceeding 5 in many positions.
Such high optical depths cause real difficulty for the present version of 
our code, since any given line of sight encounters the ``photosphere'', 
where the optical depth at some velocity is approximately unity, rather 
abruptly.  Thus, a great deal of care was taken to ensure that the solution 
was well-sampled along each line of sight. We note that the corresponding 
increase in computational time required us to decrease the spatial 
resolution of our maps.  Here, we employ a $41\times 80$ grid, as opposed 
to the $61\times 120$ grid used elsewhere. The inclination angle is 
$45^\circ$, as in the previous two cases.

The spectra shown in Fig.\ref{fig:MASSspectra} are much more complex than 
the spectra of the low mass or radiative solutions.  Many show multiple 
sharp peaks and very extended wing emission out to well past $20~km~s^{-1}$ 
(as in Shepherd \& Churchwell \cite{shch}, Gregersen et al. \cite{greg} for 
CO maps of molecular outflows, and Doeleman et al. \cite{doeleman} for SiO 
masers). We are quite certain that the multiple peaks are real, 
and not numerical artifacts, since we have verified that their location 
and appearance is independent of the spectral resolution and the degree 
to which a solution is sampled along a line of sight. It remains  
unclear to us in detail why such jagged peaks should be present, although 
it must depend on the variation of optical depth, density and velocity 
field within the `beam'. Thus behavior should be studied as a function 
of inclination to the line of sight.

Despite the unusual appearance of the spectra, the channel maps shown in 
Fig.\ref{fig:MASSmap} are quite reasonable.  A well-collimated outflow 
is apparent at velocities greater than about $10~km~s^{-1}$.  The position 
velocity diagram shown in Fig.\ref{fig:MASSpv} shows significant
emission at most positions out to approximately $20~km~s^{-1}$.  We obtain 
emission at higher velocities (up to approximately $40~km~s^{-1}$, but the 
emission is restricted to the central map positions nearest the protostar. 
The intensity-velocity diagram (Fig.\ref{fig:MASSiv}) 
again shows a power law behavior, but the 
slope is different than that obtained in the previous two cases; here we 
find that $I\propto v^{-4.1}$, which is still inside
the range allowed by the available observations 
(Cabrit et al. \cite{cab}, Richer et al. \cite{PPIV}).

%%%%%%%%%%%%%%%%%%%%%%%%%%%%%%%%%%%%%%%%%%%%%%%%%%%%%%%%%%%%%
\section{Discussion}
%%%%%%%%%%%%%%%%%%%%%%%%%%%%%%%%%%%%%%%%%%%%%%%%%%%%%%%%%%%%%

%%%%%%%%%%%%%%%%%%%%%%%%%%%%%%%%%%%%%%%%%%%%%%%%%%%%%%%%%%%%%
\subsection{The Alfv\'enic Point}
%%%%%%%%%%%%%%%%%%%%%%%%%%%%%%%%%%%%%%%%%%%%%%%%%%%%%%%%%%%%%

%%%%%%%%%%%%%%%%%%%%%%%%%%%%%%%%%%%%
\begin{figure}
%\picplace{6cm}
\psfig{figure=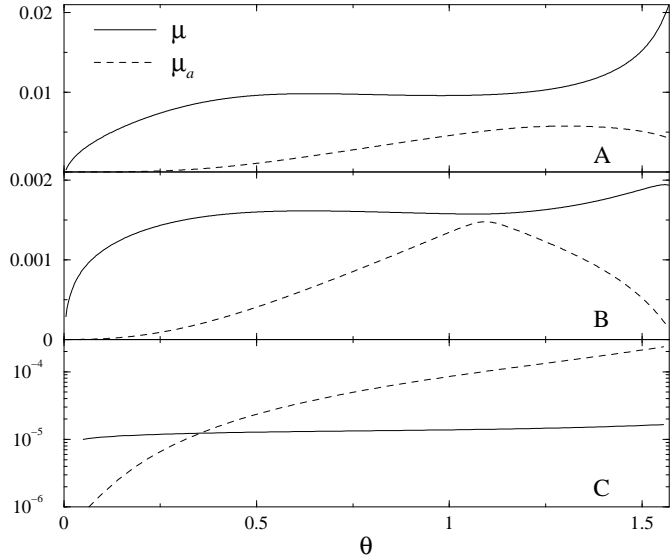,width=\linewidth}
\caption[ ]{ 
We show the density $\mu$ and the Alfv\'enic density $\mu_a$ for
really super Alfv\'enic (upper panel), slightly super Alfv\'enic 
in the region of naught radial velocity (middle panel) and critical 
(lower panel) solutions. From top to bottom Poynting flux and radial 
velocity are getting less and less important ($\alpha=-0.2$,$\Theta=0.41$). 
}
\label{figCRIT}
\end{figure}
%%%%%%%%%%%%%%%%%%%%%%%%%%%%%%%%

All the solutions presented previously were completely super-Alfv\'enic.
Here we report a trans-Alfv\'enic solution  in order to show why they 
are generally not preferred. In such a solution, the outflow size is 
reduced and the ultimate velocities are smaller. Therefore this type of 
solution is less efficient in producing high velocity outflows. Entirely 
sub-Alfv\'enic solutions can also be found {\it but they admit only pure 
accretion}. Fig.\ref{figCRIT} 
shows the density profile for two super-Alfv\'enic solutions (the two upper 
most plots A and B) and one trans-Alfv\'enic solution (C). Case (A) presents 
a model where the density is well above the critical Alfv\'enic density.
In this case, the Alfv\'enic Mach number is larger than unity. The second 
case (B) has almost the critical density  at the turning point, and 
consequently
the Alfv\'enic Mach number almost reaches unity at this point. However
the solution always remains super-Alfv\'enic. Finally the lower panel 
(case C) displays a trans-Alfv\'enic solution showing that the size of the
outflow has been reduced drastically. Therefore the most 
interesting solutions for circulation flow appear to be super-Alfv\'enic 
solutions.

%%%%%%%%%%%%%%%%%%%%%%%%%%%%%%%%%%%%%%%%%%%%%%%%%%%%%%%%%%%%%
\subsection{Comparison with FH1}
%%%%%%%%%%%%%%%%%%%%%%%%%%%%%%%%%%%%%%%%%%%%%%%%%%%%%%%%%%%%%

%%%%%%%%%%%%%%%%%%%%%%%%%%%%%%%%%%%%
\begin{figure}
%\picplace{6cm}
\psfig{figure=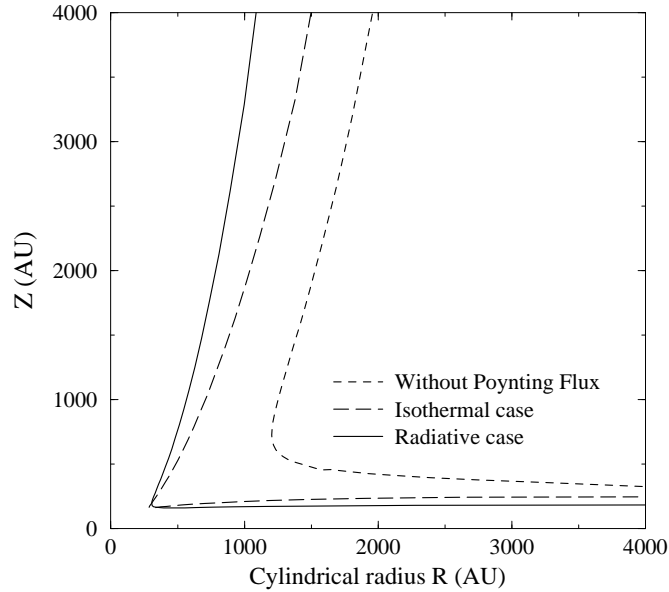,width=\linewidth}
\caption[ ]{
Comparison of streamlines without Poynting flux (dashed line)
and including Poynting flux (solid line) starting from the same 
position in the equatorial plane.
($\alpha=-0.3$,$\Theta=0.41$)
}
\label{figSTcomp}
\end{figure}
%%%%%%%%%%%%%%%%%%%%%%%%%%%%%%%%
Fiege \& Henriksen (\cite{fiege1}a, FH1) 
have shown that $r$-self-similar quadrupolar 
circulation could be a relatively realistic model for bipolar outflow.
They did not include Poynting flux that could assist the outflow as in 
the present paper. The inclusion of this effect allows us to obtain faster 
and more collimated outflows. The path followed by material 
also differs from FH1 as shown in Fig.\ref{figSTcomp}, where streamlines from 
FH1 and the present  models are plotted. All the solutions start from 
approximately the same initial position in the equatorial region with the 
same set of parameters. Streamlines from the present model get closer to 
the source than the FH1 solution, for both radiative and virial-isothermal 
cases. Thus, in the present model, the flow passes relatively close to 
the central mass. The limiting speed  is about the escape speed from the 
central mass  
(and the jet speed) of a few hundred $km~s^{-1}$. We note that the most 
collimated of the the three solutions 
is the radiative case.

%%%%%%%%%%%%%%%%%%%%%%%%%%%%%%%%%%%%%%%%%%%%%%%%%%%%%%%%%%%%%
\subsection{Stability Analysis}
%%%%%%%%%%%%%%%%%%%%%%%%%%%%%%%%%%%%%%%%%%%%%%%%%%%%%%%%%%%%%

The steady configurations presented in this article should develop strong 
shocks in the axial region  due to the strong collimation created by 
magnetic and pressure forces.  Moreover current-carrying jets, 
as in the present model, are liable to Kelvin-Helmholtz (KH), Pressure 
Driven (PD) and magnetic instabilities driven by the electrical current. 
These so called current driven (CD) instabilities have been studied only 
recently in the context of astrophysical jets (Appl \cite{appl}, Appl et 
al. \cite{LBA}). We use the results of the latter linear stability analysis 
in order to get a rough estimate of CDI and KHI in our magnetic 
configuration. Growth rates $\Gamma$ of the different modes show that
the CD helical mode ($m=1$) dominates the CD pinching mode ($m=0$) and that
CD and KH instabilities become comparable when the Mach number approaches 
unity. This is the case in the present self-similar model. The CD instability 
(with a typical wavelength $\lambda_{CD}\approx\frac{2}{5}R_{outflow}$, 
$R_{outflow}$ being the characteristic size of the outflow) should 
be located on a current sheet where reconnection and particle acceleration 
might occur. KH pinching modes ($\lambda_{KH}\approx 3 \times R_{outflow}$) 
should dominate KH helical modes, and consequently  give rise to internal
shocks in the outflow. All these results require numerical simulations
that we plan to study in another paper.

%%%%%%%%%%%%%%%%%%%%%%%%%%%%%%%%%%%%%%%%%%%%%%%%%%%%%%%%%%%%%
\section{Conclusions }
%%%%%%%%%%%%%%%%%%%%%%%%%%%%%%%%%%%%%%%%%%%%%%%%%%%%%%%%%%%%%

In this paper we have presented a model based on the $r$-self-similarity 
assumption applied to the basic equations of ideal axisymmetric and 
stationary MHD, including Poynting flux. Detailed comparisons with 
the observations have been computed  and characteristic 
scales of the problem have been given as functions of source properties. 
The luminosity needed for radiative heating is smaller even while driving 
faster outflows than in the previous model (FH1). The model geometry implies 
a natural connection between 
the fast ionized jets seen near the polar axis of the wind, and the slower 
and less-collimated molecular outflows that surround them, although 
the jets may be partly due to central activity not included 
in our model. 
  
The most massive protostars produce the fastest flows with maximum radial 
velocities in the axial region. Radiative heating produces faster 
outflows compared to the virial-isothermal case, for both low and high mass 
objects. Larger opening angles are associated with smaller magnetic 
fields. Consequently, a gradual evolutionary loss of magnetic flux may result
in outflows that widen as they age.

 Synthetic spectral lines from $^{13}CO$ (J=$1\rightarrow 0$) allow
 direct comparison with observational results via channel maps, maps of total 
emission, position-velocity and intensity-velocity diagrams.
The model reproduces well observational features. Due to internal 
instabilities in the most collimated `jet' part of the flow, 
the time evolution of the steady model should give rise to 
regularly spaced knots, and possibly excitation, ionization and 
non-thermal particle acceleration due 
to field annihilation and shock dissipation. 
 Thus the  model in its current state of development  
shows that {\it radiative heating combined with Poynting flux driving 
is efficient 
in producing high velocity outflows when the characteristic luminosity 
of the forming star (as deduced from observations) is used}. More extensive 
searches in parameter space, the detailed fit to specific sources, and 
time dependent modeling all remain to be done. Although we do not discuss 
the regions excluded from the self-similar region of the flow, the current 
correspondence 
to observational features is sufficiently exact that we expect simultaneous 
infall/outflow (`circulation') to be an essential part of any realistic 
model.

\begin{acknowledgements}
We would like to thank Sylvie Cabrit for constructive and useful discussions.
\end{acknowledgements}

\appendix
\section{The Equations \label{AP1}}
The mass flux conservation equation remains the same as in FH1:
\begin{equation}
\left (1+2\,\alpha\right )\mu\,u_{{r}}+
\frac {1}{\sin{\theta}}
\frac{d}{d\theta}\left (\mu\,u_{\theta}\sin{\theta}\right )
=0 ,
\label{MFC}
\end{equation}
while the self-similar variable $y$ is simply replaced by its 
poloidal component in the following set of equations:

\noindent
{Magnetic Flux conservation:}
\begin{equation}
{\frac {\left (\alpha+5/4\right )u_{{r}}}{y_{{p}}}}+
\frac {1}{\sin{\theta}}
{\frac {d }
{d \theta}}\left ({\frac {u_{\theta}\sin{\theta}}{y_{{p}}}}
\right )
=0
\label{BFC}
\end{equation}
\noindent
{Radial component of momentum equation:}
\begin{eqnarray}
u_{\theta}\frac {du_{r}}{d\theta}
\left (1-{M_{{ap}}}^{-2}\right )-\left ({u_{\theta}}^2+{u_
{\phi}}^2\right )\left (1-{\frac {\alpha+1/4}{M_{ap}^2}}
\right )
\nonumber 
\\
-\frac{u_{r}^2}{2}
-\left (\frac{3}{2}-2\alpha\right )\Theta+1+
\frac {u_{r}u_{\theta}}{M_{ap}^2y_{p}}\frac {dy_{p}}{d\theta}=0
\end{eqnarray}
Other equations of the system contain terms with mixed 
toroidal and poloidal components; 
\noindent
{ Angular Momentum conservation:}
\begin{eqnarray}
\frac{1}{u_{\phi}}
{\frac {du_{\phi}}{d\theta}}
\left (1-\frac {1}{M_{ap}M_{a\phi}}\right )
+\frac{u_r}{u_{\theta}}
\left (1/2-{\frac{\alpha+1/4}{M_{ap}M_{a\phi}}}\right)
\nonumber 
\\
+\cot~\theta
\left (1-{\frac {1}{M_{{ap}}M_{{a\phi}}}}\right )+{
\frac {1}
{y_{{\phi}}M_{{ap}}M_{{a\phi}}}}
{\frac {dy_{\phi}}{d\theta}}
=0
\end{eqnarray}
\noindent
{Faraday's law plus zero comoving electric field:}
\begin{equation}
{\frac {d (u_{{\phi}}u_{{\theta}})}{d \theta}}
+\left [\alpha-\frac{1}{4}\right ]u_{{\phi}}u_{{r}}+u_{{\phi}}u_{{
\theta}}{\frac {d }{d \theta}}\ln \left[\frac{1}{y_{{p}}}-\frac{1}{y_{{
\phi}}}\right]=0
\label{Faraday}
\end{equation}
\noindent
{$\theta$-component of momentum equation:}
\begin{eqnarray}
\frac{u_{r}u_{\theta}}{2}
\left (1-{\frac {2\,\alpha+1/2}{{M_{ap}}^2
}}\right )
+{u_{\phi}}^2 \cot~\theta
\left ({M_{a\phi}}^{-2}-1\right )
&
\nonumber 
\\
+{\frac {u_{r}}{{M_{ap}}^2}}{\frac {du_{r}}{d\theta}}
+u_{\theta}{\frac {du_{\theta}}{d\theta}}
+{\frac {u_\phi}{M_{ap}^2}}{\frac {du_\phi}{d\theta}}
-{\frac 
{{u_r}^2}{{y_p M_{ap}}^2}}{\frac {dy_{p}}{d\theta}}
&
\nonumber 
\\
-{\frac {
{u_\phi}^2}{{y_\phi M_{a\phi}}^2}}{\frac {dy_\phi}{d\theta}}
+{\frac {d\Theta}{d\theta}}
+{\frac {\Theta}{\mu}}{\frac {d\mu}{d\theta}}
=0&
\end{eqnarray}
\noindent
To treat the radiative heating we either hold $\Theta$ constant 
(virial-isothermal case) as discussed above, or we include the 
equations of radiative diffusion (radiative case) as in FH1. The 
equations (17), (18) and (22) of FH1 continue to apply provided 
that the optical depth given above is $\ge 2/3$. 

\section{The First integrals \label{AP2}}

Equations (\ref{MFC}) and (\ref{BFC}) together yield the first integral
\begin{equation}
\mu (u_\theta \sin{\theta})^{\frac{1/4-\alpha}{5/4+\alpha}}
y_p^{\frac{1+2\alpha}{5/4+\alpha}} \equiv q_1/4\pi,
\label{integral1}
\end{equation}
where $y_p$ replaces $y$ in FH1. Moreover directly from Eq.~\ref{MFC} 
and the poloidal stream-line equation $dr/rd\theta=u_r/u_\theta$ we have the 
stream-line integral    
\begin{equation}
r^{(1+2\alpha)}\mu u_\theta \sin{\theta} \equiv \eta_1.
\label {stream1}
\end{equation}
From this we see immediately that $\alpha>-1/2$ if we are to have quadrupolar 
stream-lines for which $r~\rightarrow~\infty$ as $u_\theta~\rightarrow~0$. A 
second stream-line integral follows from Eq.~\ref{BFC} and the stream-line 
equation, but it is not independent of Eqs.~\ref{integral1} and 
\ref{stream1} since it follows by eliminating $\mu$ between these two. The 
limit $\alpha<1/4$ follows immediately from Eq.~\ref{integral1} if we 
require finite $\mu$, $u_\theta=0$ and odd symmetry in the magnetic field at 
the equator. The integral then implies implies that $y_p\rightarrow \infty$ 
and so the field passes through zero in the equatorial plane. This conclusion 
is avoided on the polar axis since the radial velocity is free to become 
very large there. This choice has the additional merit that the equator is 
super Alfv\'enic as well as the axis. 

Using  Eq.~\ref{BFC} together with Faraday's law (\ref{Faraday}) we 
obtain the integral
\begin{equation}
u_\phi\left[\frac{1}{y_p}-\frac{1}{y_\phi}\right]
\left(\frac{\sin{\theta}}{y_p}\right)^
{\frac{1/4-\alpha}{5/4+\alpha}}
u_\theta^{\frac{3}{2(5/4+\alpha)}}=q_2.
\label{integral2}
\end{equation}
Note that the constant $q_2$ is very interesting because it can be used to 
measure the strength of the electric field which is in fact 
 \begin{equation}
\vec{E}=-\left[\frac{G M}{c r_o}\right]
\sqrt{\frac{M}{r_o^3}}
\left(\frac{r}{r_o}\right)^{\alpha-\frac{5}{4}}
\left[\frac{1}{y_p}-\frac{1}{y_\phi}\right]
\left(\vec{u_\phi}\times\vec{u_p}\right).
\end{equation}
The integral is not present in FH1 since the electric field is everywhere 
zero there.

In general we can not find an angular momentum integral explicitly (this is 
related to the difficulty of expressing scale invariance in an action 
principle; B. Gaffet, private communication) but in the special case of 
similarity index $\alpha=1/4$ such an explicit integral exists in the form 
(cf FH1): 
\begin{equation}
\frac{ u_\theta \sin{\theta}}{M_{ap}^2}\equiv q_3
\left(u_\phi \sin{\theta}\right)^3
\left(1-\frac{1}{M_{ap}M_{a\phi}}\right)^3.
\label{integral3}
\end{equation}
Although this is a very special case {\it not} normally related to our 
numerical  solutions (the mass density is constant in radius so that 
eventually the  dominance of the central mass is broken as $r$ increases), 
it is instructive to consider it further. Eq.~\ref{integral1} 
becomes simply $\mu=(q_1/4\pi)y_p^{-1}$ so that $M_{ap}^2=q_1 y_p$ and 
$M_{a\phi}^2=q_1y_\phi^2 /y_p$. Then Eqs.~\ref{integral2} and 
\ref{integral3} can be solved together  for $y_p$ and $y_\phi$ to give 
 \begin{equation}
 \frac{1}{y_p}=\frac{q_2q_3\sin^2{\theta}+q_1q_3u_\phi u_\theta 
 \sin^2{\theta}}{q_3u_\phi u_\theta \sin^2{\theta}+u_\theta^2/u_\phi^2},
 \end{equation}
 and 
 \begin{equation}
 \frac{1}{y_\phi}=\frac{q_1q_3u_\phi^4\sin^2{\theta}-q_2}
 {q_3u_\phi^4\sin^2{\theta}+u_\phi u_\theta}.
 \end{equation} 
 We therefore observe that near the equator $y_p$ can be infinite {\it only} 
 if $u_\phi\rightarrow 0$ faster than $u_\theta$. This feature is shared with 
 our numerical solutions, but of course the equator is strictly outside the 
 domain of the solution.  From the second equation we see that $y_\phi$ tends 
 to zero at the equator under the same conditions. If one were to insist that 
 $u_\phi\ne 0$ at the equator, then Eq.~\ref{integral3} shows that
 we need $M_{a\phi}M_{ap}=1$ there, or by the above Mach number definitions 
 $y_\phi=1/q_1$. Our solutions for $y_p$ and $y_\phi$ show that this is only 
 possible if $q_2=0$, that is with zero Poynting flux. Thus it seems from our
 analysis of this special case (however we always find $u_\phi=0$ at the 
 equator in our solutions) that the presence of a non-zero Poynting flux is 
 not compatible with a Keplerian equatorial disc. The material there is 
 falling radially towards the star. This is a possible non-linear end state for
 an instability that couples disc rotation to Alfv\'en waves propagating out
 of the plane (Shu et al. \cite{shu94}, Dendy et al. \cite{dentagger}).

\end{document}